\documentstyle[epsf,12pt]{article}
\newcommand {\eq}{\begin{equation}}
\newcommand {\qe}{\end{equation}}
\newcommand {\cen}[1]{\begin{center} #1 \end{center}}
\newcommand {\bfr}{{\bf r}}

\newcommand {\ea} {{\it et al.}}
\newcommand {\epp}{\epsilon^+}
\newcommand {\epm}{\epsilon^-}
\newcommand {\epz}{\epsilon^0}
\newcommand {\bfk}{{\bf k}}
\newcommand {\bfp}{{\bf p}}
\newcommand {\prc}{Phys. Rev. C}

\newcommand {\th}{{\tiny \frac{1}{2}}}
\newcommand {\rti}{\frac{1}{\sqrt{2}}}
\newcommand {\calo}{{\cal O}}

\newcommand {\bfsigma} {\mbox {\boldmath $\sigma$}}

\newcommand {\bfeps} {\mbox {\boldmath $\epsilon$}}
\newcommand {\pim} {$\pi^-$}

\newcommand {\pr}{{\it Phys. Rev. }}
\newcommand {\prl}{{\it Phys. Rev. Lett. }}
\newcommand {\pl}{{\it Phys. Lett. }}
\newcommand {\nucp}{{\it Nucl. Phys. }}
\begin{document}
\baselineskip=1.2\baselineskip


\begin{center}

{\Huge $\Lambda$-Neutron Scattering Lengths from Radiative 
K$^-$ Capture}

\vspace{.5in}

{\large W. R. Gibbs, S. A. Coon, H. K. Han}

\vspace{.15in}

Department of Physics, New Mexico State University, Las Cruces, NM, 88003

\vspace{.3in}

{\large B. F. Gibson}

\vspace{.15in}

Theoretical Division, Los Alamos National Laboratory, Los Alamos, NM, 87545

\vspace{.3in}

\today
\end{center}

\vspace{.5in}

\abstract{Radiative capture of the K$^-$ by the deuteron is 
examined as a reaction for measurement of the $\Lambda$-neutron
scattering lengths. Using the final state interaction,
analogous to measurements of the radiative capture
of pions by deuterium, the scattering lengths of both the triplet
and singlet $\Lambda$-n interaction can be inferred. The problem
of the separation of these two fundamental parameters without
and with spin information in the experiment
is addressed. It is shown that a measurement of the deviation
of the vector deuteron polarization asymmetry from --1 provides
a direct determination of the difference of the singlet and
triplet scattering lengths.}

\newpage

\section{Introduction}

The interaction of the nucleon with the $\Lambda$ hyperon
is of great theoretical interest. Since the 
one-pion-exchange interaction is absent (except under broken
isospin \cite{pruhonice}), one is able to focus
on the shorter range contributions to the interaction.  The 
singlet and triplet scattering lengths are predicted to be
equal by SU(6) symmetry. 

Unfortunately,
experiments are difficult to perform for this system because of the 
short lifetime of the $\Lambda$.  Nevertheless, some experimental results
do exist\cite{alexander} and have been analyzed\cite{rijken,reuber}.
Rijken \ea \cite{rijken} have recently generated a set of potentials 
which fit these data.  The range of scattering lengths obtained from 
this analysis is rather large, which is perhaps a natural consequence 
considering that the data do not extend to low energies.

Use of the radiative capture of a \pim\ from an s-wave 
atomic state by deuterium to measure the neutron-neutron (nn) 
scattering length, through the influence of the final state 
interaction, was investigated in an experiment by Phillips and 
Crowe\cite{crowe} and soon after studied theoretically 
 by McVoy\cite{mcvoy} and Bander\cite{bander}.
The basic technique consists of comparing the shape of the 
of the photon spectrum with calculations.
Because of the final-state interaction of the two neutrons,
a peak results near the high-energy end of the spectrum 
corresponding to a maximum in the s-wave phase shifts at low 
neutron-neutron energy. Since this phase shift rises very 
quickly to a maximum, as a function of energy in the neutron-neutron 
frame, the peak in the photon spectrum is very near the maximum 
energy. It is also possible to use the neutron spectrum at low 
energies for the studies provided that the angle between the neutron 
and photon is known.  Experiments have been performed in both 
geometries.

These early exploratory efforts were followed by an experiment 
detecting the neutrons\cite{hadock} which gave useful results.
Following a study of limits due to the theoretical 
uncertainties\cite{ggs}, experiments were carried out at
the Paul Scherrer Institute (PSI) detecting only the photon, as 
well as experiments in which the photon and neutron were detected 
in coincidence\cite{psi}. These experiments resulted in values
for the nn scattering length (and effective range) which
were considered to be the best obtained to date.  Recently 
another experiment was performed using this reaction at the Clinton 
P. Anderson Meson Physics Facility\cite{torrow}. Results 
from this experiment appear to confirm the PSI results. The
uncertainty in the scattering length is of the order of $\pm 0.5$ 
fm for a value of -18.58 fm, about 2-3\%.

A determination of the $\Lambda n$ scattering lengths from
the photon spectrum of the reaction 
K$^-$d$\rightarrow$n$\Lambda \gamma$, where the capture takes 
place from an atomic state, was suggested by Gibson \ea 
\cite{gibson}. A study of this 
process performed by Workman and Fearing \cite{workman}  
concluded that the different hadronic routes considered by Akhiezer 
\ea \cite{akhiezer} had a negligible impact on a possible 
measurement (however, see Refs. \cite{cotanch} for comments on
their representation of the amplitudes).  They also concluded that 
the dominant operator for the conversion of a proton to a $\Lambda$ 
through radiative capture of a kaon was of the Kroll-Ruderman 
form

\eq {\cal O}=\bfsigma\cdot\bfeps \qe
where $\bfeps$ is the polarization of the outgoing photon
and $\bfsigma$ is the spin operator for the proton on which
the capture takes place.

At the same time as this theoretical effort, a 
feasibility study\cite{gall} demonstrated 
that the spectrum for this reaction could be separated from the 
background.  Workman and Fearing compared their analysis with
these data to show that reasonable values of the scattering lengths
were consistent with the data.

Two important issues were left unresolved by the work of
Workman and Fearing.  They used
only the asymptotic $\Lambda$-neutron wave functions. In
the work of Ref. \cite{ggs} the lack of knowledge of the short
range final-state wave function was found to be the dominant 
theoretical uncertainty in the analysis of the scattering
length, of the order of $\pm 0.3$ fm.  If the uncertainty is 
of the same order in the case of $\Lambda$-neutron scattering 
the consequences are more serious since the scattering length 
itself is thought to be an order of magnitude smaller.

A second problem, which may be more serious, is that,
while in the case of the neutron-neutron final state the
spin state of the two neutrons is restricted to be singlet
by the Pauli principle, the $\Lambda$n final state
consists of a mixture of a singlet and triplet states.  
One expects that the scattering in these two states may
be different, and indeed the dependence of the scattering
length on the spin might provide some very important clues
to the structure of the interaction.  While it has been
suspected that polarization information may provide a
tool to obtain a singlet-triplet separation
(see Balewski \ea \cite{balewski} for example), we provide
here the formalism and expressions necessary to obtain
such a separation for this reaction.

We first give a brief heuristic description
of the basic physics underlying the development to follow.
Consider a deuteron target prepared in a configuration with only
magnetic spin projection zero along a z-axis defined by the 
direction of the photon.  Assume also that the final state of the 
$\Lambda$-n system is in a singlet state.  Since the $K^-$ is 
assumed to be captured from an atomic s-state, there will be a 
total spin projection of zero in the initial state and zero
angular momentum projection in the final state, other than that due 
to the spin of the $\gamma$. Since the ``transversality'' relation,

\eq
\bfeps\cdot\bfk =0,
\label{trans}\qe
requires that the spin projection of the photon must lie along 
its direction of travel (i.e. there is no zero spin projection) 
this assumed transition to the singlet final state must have
zero amplitude due to the conservation of the z-projection of 
angular momentum. Thus, for example, a spectrum taken under 
these conditions could be analyzed to obtain the 
triplet scattering parameters alone.

The previous paragraph gives only an example of how deuteron 
spin information
can affect the analysis.  A measurement of the photon circular
polarization can also help to separate the singlet and triplet
scattering states. 
In the subsequent sections we develop the expressions necessary
to calculate the photon spectrum and provide specific results.

\section{Expressions for the Rate}

For comparison with the measurement we need the theoretical
prediction for the shape of the spectrum.  The absolute rate
is not usually measured in this type of experiment. 

\subsection{Kinematics and Phase Space}

The magnitude of the momentum, $\bfp$, of each baryon in this
pair's center of mass is given by 
\eq p^2=\frac{(s+M_n^2-M_{\Lambda}^2)^2}{4s}-M_n^2 \qe
with $s$ the center of mass energy of the $\Lambda$-neutron
pair given by
\eq s=E_0^2-2kE_0, 
\qe
where $E_0$ is the total energy of the initial system and $k$
is the final photon momentum.  $M_{\Lambda}$
and $M_n$ are the masses of the $\Lambda$ and neutron respectively.
 
Since we shall assume that the final-state baryons are not observed, 
we must integrate over their directions, giving for the differential
rate of the observed photon
\eq 
\frac{d \Gamma}{d\Omega_{\bfk}dk}\propto \frac{pk\omega_{\Lambda}
\omega_n}{(\omega_{\Lambda}+\omega_n)}\int |M|^2 d\Omega_{\bfp}
\qe
The total energies, $\omega_n$ and $\omega_{\Lambda}$ are 
very nearly constant over the energy range of interest.

It should be noted that a single factor of the photon momentum
$k$ appears.  It is linear only with relativistic phase space;
it would appear to the second power in non-relativistic phase 
space. When this reaction was first proposed, the non-relativistic
version was usually used (in fact Williams \cite{williams}
used this reaction as an example of the use of non-relativistic
phase space).  Of course, either type of phase space can
be used provided that the proper corresponding operator is used.  
The difference between a first and second power of the photon
momentum  
is relativistic in origin and the effect is very small in many
cases.  In the analysis of Ref. \cite{psi} only the very
end of the spectrum was used and the photon momentum is
sufficiently constant over this range that such a factor
is immaterial.  For the analysis of Ref. \cite{torrow}
the larger range of the spectrum used leads to a small
sensitivity to the power of $k$ used in this expression.
For the present reaction, since the peak lies farther
from the end point of the spectrum than for the 
$\pi^-d\rightarrow nn\gamma$ reaction, a larger range of photon 
momenta may be included in the analysis and the correct factor 
is more important.

\subsection{Matrix Element}

The mathematical development of the expressions for the matrix
element used here 
follows a different procedure than that of Refs. \cite{ggs}
and \cite{workman}.  The previous methods have written the
matrix element for the transition first as a plane wave transition
then corrected this expression by adding and subtracting the
s-wave contribution.  The s-wave contribution with a final
state interaction is then substituted in the added element.
For a matrix element
with two particles detected in the final state this is the
appropriate method.  However, for the case in which only
the photon is detected, there is an average over the unobserved
baryon momenta which must be made as outlined in the previous
section.  This integral was carried out numerically in the 
previous developments.

In the method used in the present analysis the partial-wave 
quantities are calculated in terms of the relative 
$\Lambda$-neutron momentum.  This technique 
requires several partial waves to be computed for a handful 
of magnetic quantum numbers.  While it may seem
{\it a priori} less efficient, in the end it has several
advantages.  The first is that the final averaging over the
direction of the unobserved momenta can be done analytically.  Since the 
contributions from the higher partial waves depend only on the initial 
state wave functions (we assume no interaction in partial waves
higher than $\ell=0$), if one wishes to fit the data by varying
the scattering length and effective range for the singlet and
triplet states, these higher partial waves need be calculated only once.

A second advantage is that the degree of coherence or incoherence
of the singlet and triplet states is manifest.  For example, we will 
see that if all magnetic quantum numbers of the initial deuteron
are equally populated and the polarization of the final photon
is not observed, the transitions to the singlet and triplet
final states are incoherent.

The matrix element for the reaction which proceeds from an
initial deuteron with spin-projection $S_z$ to a final state
of the $\Lambda$-neutron system in a total spin state 
$(S', S_z')$ is
\eq
M^{S',S_z',S_z}=\int d\bfr_1d\bfr_2
<S'S_z'|\psi^-_{S'}(\bfp,\bfr)e^{-i{\bf K}
\cdot(\bfr_1+\bfr_2)}|\calo e^{i\bfk\cdot\bfr_1}|\Psi_D^{S_z}(\bfr)>
\qe
where $\psi^-_{S'}(\bfp,\bfr)$ is the final state wave function
of the neutron-$\Lambda$ system and the variable $S'$ is 0
or 1 for the singlet or triplet case. ${\bf K}$ is the center-of-mass
momentum of the baryon pair. The quantity 
$|\Psi_D^{S_z}(\bfr)>$ is the deuteron wave function with an initial
spin projection $S_z$. We obtain the radial deuteron wave function from 
the solution
of the coupled Schr\"odinger equations for a pure one-pion-exchange
potential.  It has been shown that this interaction reproduces all
of the low-energy properties of the deuteron \cite{rosa,fg,ballot}
(likely to be the most important in this calculation).
The operator ${\cal O}=\bfsigma_1\cdot\bfeps$.

Transforming to relative and center-of-mass
coordinates in the spatial variables we have
\eq
M^{S',S_z',S_z}=(2\pi)^3\delta({\bf K}-\bfk)\int d\bfr
<S'S_z'|\psi^*_{S'}(\bfp,\bfr)|\calo e^{i\frac{\bfk\cdot\bfr}{2}}
|\Psi_D^{S_z}(\bfr)>.
\qe

We write the deuteron wave function as
\eq 
|\Psi_D^{S_z}(\bfr)>=S(r)Y_0^0|1S_z>+D(r)\sum Y_2^{m'}(\bfr)
|1\sigma>C_{1,\ \ 2,\ \ 1}^{\sigma,m',S_z},
\qe
and the $\Lambda$-n relative motion wave function as
\eq
\psi_{S'}(\bfp,\bfr)=4\pi\sum i^{\ell}Y_{\ell}^m(\bfr)Y_{\ell}^{m*}(\bfp)
\phi_{\ell}^{S'}(r).
\qe

With these expansions we can write (with an arbitrary overall
normalization, omitting the delta function)

$$
M^{S',S_z',S_z}=4\pi <S'S_z'|\calo |
\int d\bfr \sum_{\ell,m,L,M} i^{L-\ell}Y_{\ell}^{m*}(\bfr)
Y_L^{M}(\bfr)Y_{\ell}^{m}(\bfp)Y_L^{M*}(\bfk)
\phi_{\ell}^{S'}(r)j_L\left(\frac{kr}{2}\right)$$

\eq
\times\left[S(r)Y_0^0|1S_z>+D(r)\sum Y_2^{m'}(\bfr)
|1\sigma>C_{1,\ \ 2,\ \ 1}^{\sigma,m',S_z}\right].
\qe
With the photon direction, $\bfk$, along the z axis, one has

\eq
Y_L^M(\bfk)=\delta_{M,0}\sqrt{\frac{2L+1}{4\pi}}.
\qe
Performing the integral over the angles of $\bfr$, we obtain

$$
M^{S',S_z',S_z}= \sum_{\ell,m}Y_{\ell}^m(\bfp)[<S'S_z'|\calo |1S_z>
\delta_{m0}
\sqrt{2\ell+1}\int_0^{\infty} r^2dr \phi_{\ell}^{S'}(r)j_{\ell}(\th kr)S(r)
$$

$$
+\sum_{L,\sigma}i^{L-\ell}
\sqrt{\frac{5(2L+1)}{(2\ell+1)}}
C_{L,2,\ \ell}^{0,m,m}C_{L,2,\ell}^{0,0,0}
\sqrt{2L+1}C_{1,\ 2,\ 1}^{\sigma,m,S_z}<S'S_z'|\calo |1\sigma> $$

\eq \times
\int_0^{\infty} r^2dr D(r)\phi_{\ell}^{S'}(r)j_L\left(\frac{kr}{2}\right)]
\qe
or,

$$M^{S',S_z',S_z}
=\sum_{\ell,m}Y_{\ell}^m(\bfp)
[S_{\ell}(S')\delta_{m,0}<S'S_z'|\calo |1S_z>
$$

\eq
+\sqrt{5}\sum_{\sigma}C_{1,\ 2,\ 1}^{\sigma,m,S_z}
<S'S_z'|\calo |1\sigma>\sum_{L} i^{L-\ell}
C_{\ell\,2\ \ ,L}^{m,-m,0}C_{\ell,2,L}^{0,0,0}D_{\ell,L}(S')]
\qe
with the definitions
\eq
S_{\ell}(S')=\sqrt{2\ell+1}\int_0^{\infty} r^2dr 
\phi_{\ell}^{S'}(r)j_{\ell}(\frac{kr}{2})S(r)
\qe

\eq 
D_{\ell,L}^{S'}=\sqrt{2\ell+1}\int_0^{\infty} r^2dr 
D(r)\phi_{\ell}^{S'}(r)j_L\left(\frac{kr}{2}\right).
\qe
If we define
\eq
B_{\ell}^m(S')\equiv \sum_Li^{L-\ell}
C_{\ell,2,L}^{m,-m,0}C_{\ell\ ,2\ \ ,L}^{0,0,0}D_{\ell,L}(S'),
\qe
we can write
\eq
M^{S',S_z',S_z}=\sum_{\ell,m}Y_{\ell}^m(\bfp)
\left[S_{\ell}\delta_{m,0}<S'S_z'|\calo |1S_z>
+\sqrt{5}\sum_{\sigma}C_{1,\ 2,\ 1}^{\sigma,m,S_z}
<S'S_z'|\calo |1\sigma>B_{\ell}^m\right],
\qe
where the explicit dependence on $S'$ of $S_{\ell}$ and 
$B_{\ell}^m$ has been suppressed.

We define the spherical components of the photon polarization 
vector as
\eq
\epsilon^{+1}=-\frac{1}{\sqrt{2}}(\epsilon_x+i\epsilon_y);\ \ 
\epsilon^{-1}=\frac{1}{\sqrt{2}}(\epsilon_x-i\epsilon_y);\ \ 
\epsilon^0=\epsilon_z .
\label{spindefs}\qe
The spin structure of the radiative capture operator is assumed to 
be of the form 
\eq
\calo=\bfsigma_1\cdot\bfeps
\qe
where particle 1 is the proton which is transformed into a $\Lambda$.
The expectation values of this operator for the baryon spin states 
are 

$$ \begin{array}{|l|r|}
\hline
\ \ \ \ \ {\rm Triplet}&\\
\hline
<1,0|\calo|1,1>&\epp \\
<1,0|\calo|1,-1>&-\epm \\
<1,0|\calo|1,0>&0 \\
<1,1|\calo|1,1>&\epz \\
<1,-1|\calo|1,-1>&-\epz \\
<1,1|\calo|1,0>&-\epm \\
<1,-1|\calo|1,0>&\epp \\
<1,1|\calo|1,-1>&0 \\
<1,-1|\calo|1,1>&0 \\
\hline
\ \ \ \ {\rm Singlet}&\\
\hline
<0,0|\calo|1,1>&-\epp \\
<0,0|\calo|1,-1>&-\epm \\
<0,0|\calo|1,0>&\epz \\
\hline
\end{array} $$
The transversality condition gives $\epz=0$.

Because of the averaging over the direction of the unobserved
momentum $\bfp$, each term in $\ell$ and $m$ will contribute
incoherently. We now consider each of the cases $m=0$, $m=\pm 1$ and
$m=\pm 2$ separately.  In the square-averaging over the photon
polarizations we omit the uniform factor of 1/2.

\subsubsection{m=0}

For $m=0$

\eq
M^{S',S_z',S_z}_{m=0}=<S'S'_z|\calo |1S_z>
\sum_{\ell=0}^{\infty}Y_{\ell}^0(\bfp)(S_{\ell}
+\sqrt{5}C_{1,\ \ 2,\ \ 1}^{S_z,0,S_z}B_{\ell}^0)
\qe
This term will contribute to the transitions 
$(S_z=\pm 1\rightarrow S'_z=0)$ and $(S_z=0\rightarrow S'_z=\pm 1)$ 

\eq
M_{m=0}^{1,0,-1}=-\epm
\sum_{\ell=0}^{\infty} Y_{\ell}^0(\bfp)(S_{\ell}+\rti B_{\ell}^0)
=M_{m=0}^{0,0,-1}\qe

\eq
M_{m=0}^{1,0,1}=\epp
\sum_{\ell=0}^{\infty} Y_{\ell}^0(\bfp)(S_{\ell}+\rti B_{\ell}^0)
=-M_{m=0}^{0,0,1}\qe

\eq
M_{m=0}^{1,1,0}=-\epm
\sum_{\ell=0}^{\infty} Y_{\ell}^0(\bfp)(S_{\ell}-\sqrt{2} B_{\ell}^0)
;\ \ \ M_{m=0}^{1,-1,0}=\epp
\sum_{\ell=0}^{\infty} Y_{\ell}^0(\bfp)(S_{\ell}-\sqrt{2} B_{\ell}^0)
\qe
giving a contribution to the total capture rate of
\eq
\sum_{\ell=0}^{\infty}(S_{\ell}+\rti B_{\ell}^0)^2
\qe
from each of the initial magnetic states $\pm 1$ for either the singlet
or triplet final state and 
\eq
2\sum_{\ell=0}^{\infty}(S_{\ell}-\sqrt{2} B_{\ell}^0)^2 
\qe
from the initial magnetic state $0$ for the triplet final state.

Since the singlet and triplet states add constructively for $S_z=1$
and destructively for $S_z=-1$, they are incoherent if and
only if the populations of the these two initial states are equal.

\subsubsection{$|m|=1$}

For $|m|=1$

\eq
M^{S',S_z',S_z}_{m=1}=<S'S'_z|\calo |1S_z-1>
\sqrt{5}C_{1,\ \ 2,\ \ 1}^{S_z-1,1,S_z}           
\sum_{\ell=1}^{\infty}Y_{\ell}^1(\bfp)B_{\ell}^1
\qe
contributes to $(S_z=0\rightarrow S'_z=0)$;

\eq
M_{m=1}^{1,0,0}=-\epm\sqrt{\frac{3}{2}}
\sum_{\ell=1}^{\infty} Y_{\ell}^1(\bfp)B_{\ell}^1
=M_{m=1}^{0,0,0}
\qe
and $(S_z=1\rightarrow S'_z=\pm 1)$

\eq
M_{m=1}^{1,1,1}=\epm\sqrt{\frac{3}{2}}
\sum_{\ell=1}^{\infty} Y_{\ell}^1(\bfp)B_{\ell}^1
;\ \ \ M_{m=1}^{1,-1,1}=-\epp\sqrt{\frac{3}{2}}
\sum_{\ell=1}^{\infty} Y_{\ell}^1(\bfp)B_{\ell}^1
\qe
while

\eq
M^{S',S'_z,S_z}_{m=-1}=<S'S'_z|\calo |1S_z+1>
\sqrt{5}C_{1,\ \ 2,\ \ 1}^{S_z+1,-1,S_z}           
\sum_{\ell=1}^{\infty}Y_{\ell}^{-1}(\bfp)B_{\ell}^1
\qe
contributes to $(S_z=0\rightarrow S'_z=0)$;

\eq
M_{m=-1}^{1,0,0}=\epp\sqrt{\frac{3}{2}}
\sum_{\ell=1}^{\infty} Y_{\ell}^{-1}(\bfp)B_{\ell}^1
=-M_{m=-1}^{0,0,0}
\qe
and $(S_z=-1\rightarrow S'_z=\pm 1)$

\eq
M_{m=-1}^{1,1,-1}=\epm\sqrt{\frac{3}{2}}
\sum_{\ell=1}^{\infty} Y_{\ell}^{-1}(\bfp)B_{\ell}^1
;\ \ \ M_{m=-1}^{1,-1,-1}=-\epp\sqrt{\frac{3}{2}}
\sum_{\ell=1}^{\infty} Y_{\ell}^{-1}(\bfp)B_{\ell}^1
\qe
giving a contribution from $|m|=1$ to the capture rate of
\eq 
3\sum_{\ell=1}^{\infty}\left(B_{\ell}^1\right)^2;\ \ 
{\rm for\ each\ of\ }\ S_z=0, \pm 1.
\qe
Only the $S_z=0$ state contributes for the singlet final state.
Since the relative sign of the singlet and triplet is different
for $m=1$ and $m=-1$, the singlet and triplet states are always
incoherent for $|m|=1$.

\subsubsection{$|m|=2$}

For $|m|=2$

\eq
M^{S',S'_z,S_z}_{m=2}=<S'S'_z|\calo |1S_z-2>
\sqrt{5}C_{1,\ \ 2,\ \ 1}^{S_z-2,2,S_z}           
\sum_{\ell=2}^{\infty}Y_{\ell}^2(\bfp)B_{\ell}^2
\qe
contributes to $(S_z=1\rightarrow S'_z=0)$

\eq
M_{m=2}^{1,0,1}=-\epm\sqrt{3}
\sum_{\ell=2}^{\infty} Y_{\ell}^2(\bfp)B_{\ell}^2
=M_{m=2}^{0,0,1}
\qe
while

\eq
M^{S',S'_z,S_z}_{m=-2}=<S'S'_z|\calo |1S_z+2>
\sqrt{5}C_{1,\ \ 2,\ \ 1}^{S_z+2,-2,S_z}           
\sum_{\ell=2}^{\infty}Y_{\ell}^{-2}(\bfp)B_{\ell}^2
\qe
contributes to $(S_z=-1\rightarrow S'_z=0)$.

\eq
M_{m=-2}^{1,0,-1}=\epp\sqrt{3}
\sum_{\ell=2}^{\infty} Y_{\ell}^{-2}(\bfp)B_{\ell}^2
=-M_{m=-2}^{0,0,-1}
\qe
giving a contribution to the capture rate for either initial spin
projection $\pm 1$ of
\eq
3\sum_{\ell=2}^{\infty}\left(B_{\ell}^2\right)^2
\qe

Since the singlet and triplet states add constructively for $S_z=1$
and destructively for $S_z=-1$, they are incoherent only if the 
 populations of the these two initial states are equal.

\subsection{Summary}

Combining the expressions in the previous section we obtain the following
results for the capture rate for the various initial and final states.
The notation for quantities calculated with the singlet or triplet final
state on the $\Lambda$-n system is now shown explicitly.
For the amplitudes in the form $M^{S'_z,S_z}_m$

\cen{$m=0, \ \epm$}

\eq 
M^{0,-1}_0=-\sum_{\ell=0}^{\infty}Y_{\ell}^0(\bfp)\left[
S_{\ell}(1)+\frac{1}{\sqrt{2}}B_{\ell}^0(1)+
S_{\ell}(0)+\frac{1}{\sqrt{2}}B_{\ell}^0(0)\right]
\qe
\eq
M^{1,0}_0=-\sum_{\ell=0}^{\infty}Y_{\ell}^0(\bfp)
\left(S_{\ell}(1)-\sqrt{2}B_{\ell}^0(1)\right)
\qe

\cen{$m=0, \ \epp$}

\eq 
M^{0,1}_0=\sum_{\ell=0}^{\infty}Y_{\ell}^0(\bfp)\left[
S_{\ell}(1)+\frac{1}{\sqrt{2}}B_{\ell}^0(1)-
S_{\ell}(0)-\frac{1}{\sqrt{2}}B_{\ell}^0(0)\right]
\qe
\eq
M^{-1,0}_0=\sum_{\ell=0}^{\infty}Y_{\ell}^0(\bfp)
\left(S_{\ell}(1)-\sqrt{2}B_{\ell}^0(1)\right)
\qe

\cen{$m=1, \ \epm$}
\eq
M^{0,0}_1=-\sqrt{\frac{3}{2}}\sum Y_{\ell}^1(\bfp)
\left[B_{\ell}^1(1)+B_{\ell}^1(0)\right]
\qe

\eq
M^{1,1}_1=\sqrt{\frac{3}{2}}\sum Y_{\ell}^1(\bfp)B_{\ell}^1(1)
\qe

\cen{$m=1, \ \epp$}
\eq
M^{-1,1}_1=-\sqrt{\frac{3}{2}}\sum Y_{\ell}^1(\bfp)B_{\ell}^1(1)
\qe

\cen{$m=-1, \ \epm$}

\eq
M^{1,-1}_{-1}=\sqrt{\frac{3}{2}}\sum_{\ell=0}^{\infty}
Y_{\ell}^{-1}(\bfp)B_{\ell}^0(1)
\qe

\cen{$m=-1, \ \epp$}

\eq
M^{0,0}_{-1}=\sqrt{\frac{3}{2}}\sum Y_{\ell}^{-1}1(\bfp)
\left[B_{\ell}^1(1)-B_{\ell}^1(0)\right]
\qe

\eq
M^{-1,-1}_{-1}=\sqrt{\frac{3}{2}}\sum Y_{\ell}^{-1}(\bfp)B_{\ell}^1(1)
\qe

\cen{$m=2, \ \epm$}

\eq
M^{0,1}_{2}=-\sqrt{3}\sum Y_{\ell}^2(\bfp)
\left[B_{\ell}^1(1)+B_{\ell}^1(0)\right]
\qe

\cen{$m=-2, \ \epp$} 

\eq
M^{0,-1}_{-2}=\sqrt{3}\sum Y_{\ell}^{-2}(\bfp)
\left[B_{\ell}^1(1)-B_{\ell}^1(0)\right].
\qe

For the capture rates we have

$$
S_z=1;\ \ \sum_{\ell=0}^{\infty}\left[S_{\ell}(1)+
\rti B_{\ell}^0(1) -S_{\ell}(0)-\rti B_{\ell}^0(0)\right]^2+
3\sum_{\ell=1}^{\infty}\left[B_{\ell}^1(1)\right]^2
$$

\eq
+3\sum_{\ell=2}^{\infty}\left[B_{\ell}^2(1)+B_{\ell}^2(0)\right]^2
\qe

\eq 
S_z=0;\ \ 2\sum_{\ell=0}^{\infty}\left[S_{\ell}(1)-
\sqrt{2} B_{\ell}^0(1)\right]^2+
3\sum_{\ell=1}^{\infty}\left[B_{\ell}^1(1)\right]^2
+3\sum_{\ell=1}^{\infty}\left[B_{\ell}^1(0)\right]^2\ \ \ \ \ \ \ 
\ \ \ \ \ \ 
\qe

$$
S_z=-1;\ \ \sum_{\ell=0}^{\infty}\left[S_{\ell}(1)+
\rti B_{\ell}^0(1)+S_{\ell}(0)+\rti B_{\ell}^0(0)\right]^2+
3\sum_{\ell=1}^{\infty}\left[B_{\ell}^1(1)\right]^2
$$

\eq
+3\sum_{\ell=2}^{\infty}\left[B_{\ell}^2(1)-B_{\ell}^2(0)\right]^2.
\qe
Using plane waves for $\ell \ge 1$

\eq 
S_z=1;\ \ \ 
\left\{S_0(1)-S_0(0)+\rti \left[B_0^0(1)-B_0^0(0)\right]\right\}^2+
3\sum_{\ell=1}^{\infty}\left[B_{\ell}^1\right]^2
+12\sum_{\ell=2}^{\infty}\left[B_{\ell}^2\right]^2\ \ \ \ 
\qe

\eq 
S_z=0;\ \ 
2\left[S_0(1)-\sqrt{2} B_0^0(1)\right]^2+
2\sum_{\ell=1}^{\infty}\left[S_{\ell}(1)-\sqrt{2} B_{\ell}^0(1)\right]^2
+6\sum_{\ell=1}^{\infty}\left[B_{\ell}^1\right]^2\ \ \ \ \ \ \ \ \ \ 
\ \ \ 
\qe

\eq
S_z=-1;\ \ 
\left\{S_0(1)+S_0(0)+\rti \left[B_0^0(1)+B_0^0(0)\right]\right\}^2+
4\sum_{\ell=1}^{\infty}\left[S_{\ell}+\rti B_{\ell}^0\right]^2+
3\sum_{\ell=1}^{\infty}\left[B_{\ell}^1\right]^2
\qe
Adding, and assuming all magnetic projections of the deuteron to be
 equally populated, we find

$$
2\left[S_0(1)+\rti B_0^0(1)\right]^2+2\left[S_0(0)+\rti B_0^0(0)\right]^2+
2\left[S_0(1)-\sqrt{2} B_0^0(1)\right]^2
$$
\eq
+6\sum_{\ell=1}^{\infty}\left\{
\left[S_{\ell}\right]^2+\left[B_{\ell}^0\right]^2\right\}+
12\sum_{\ell=1}^{\infty}\left[B_{\ell}^1\right]^2+
12\sum_{\ell=2}^{\infty}\left[B_{\ell}^2\right]^2.
\qe

Also useful are the expressions for a given photon polarization.

\cen{$\epm$}

\eq 
S_z=1;\ \ \ 
\frac{3}{2}\sum_{\ell=1}^{\infty}\left[B_{\ell}^1\right]^2
+12\sum_{\ell=2}^{\infty}\left[B_{\ell}^2\right]^2\ \ \ \ 
\qe

\eq 
S_z=0;\ \ 
\left|S_0(1)-\sqrt{2} B_0^0(1)\right|^2+
\sum_{\ell=1}^{\infty}\left[S_{\ell}(1)-\sqrt{2} B_{\ell}^0(1)\right]^2
+6\sum_{\ell=1}^{\infty}\left[B_{\ell}^1\right]^2\ \ \ \ \ \ \ \ \ \ 
\ \ \ 
\qe

\eq
S_z=-1;\ \ 
\left|S_0(1)+S_0(0)+\rti \left[B_0^0(1)+B_0^0(0)\right]\right|^2+
4\sum_{\ell=1}^{\infty}\left[S_{\ell}+\rti B_{\ell}^0\right]^2+
\frac{3}{2}\sum_{\ell=1}^{\infty}\left[B_{\ell}^1\right]^2
\qe

\cen{$\epp$}

\eq 
S_z=1;\ \ \ 
\left|S_0(1)-S_0(0)+\rti \left[B_0^0(1)-B_0^0(0)\right]\right|^2+
\frac{3}{2}\sum_{\ell=1}^{\infty}\left[B_{\ell}^1\right]^2 
\qe

\eq 
S_z=0;\ \ 
\left|S_0(1)-\sqrt{2} B_0^0(1)\right|^2+
\sum_{\ell=1}^{\infty}\left[S_{\ell}(1)-\sqrt{2} B_{\ell}^0(1)\right]^2
\qe

\eq
S_z=-1;\ \ \frac{3}{2}\sum_{\ell=1}^{\infty}\left[B_{\ell}^1\right]^2.
\qe

\subsection{Discussion}

The ratio of the triplet to singlet rates, in the limit of s-wave 
contributions and an s-wave deuteron only, and assuming equal
singlet and triplet scattering in the s-wave, is two. The
introduction of the higher partial waves, and a difference
in the scattering lengths, modifies this ratio but it is still
roughly two.  Hence there is a greater sensitivity to the triplet
scattering length than to the singlet.

\begin{figure}[htb]
\epsfysize=180mm
\epsffile{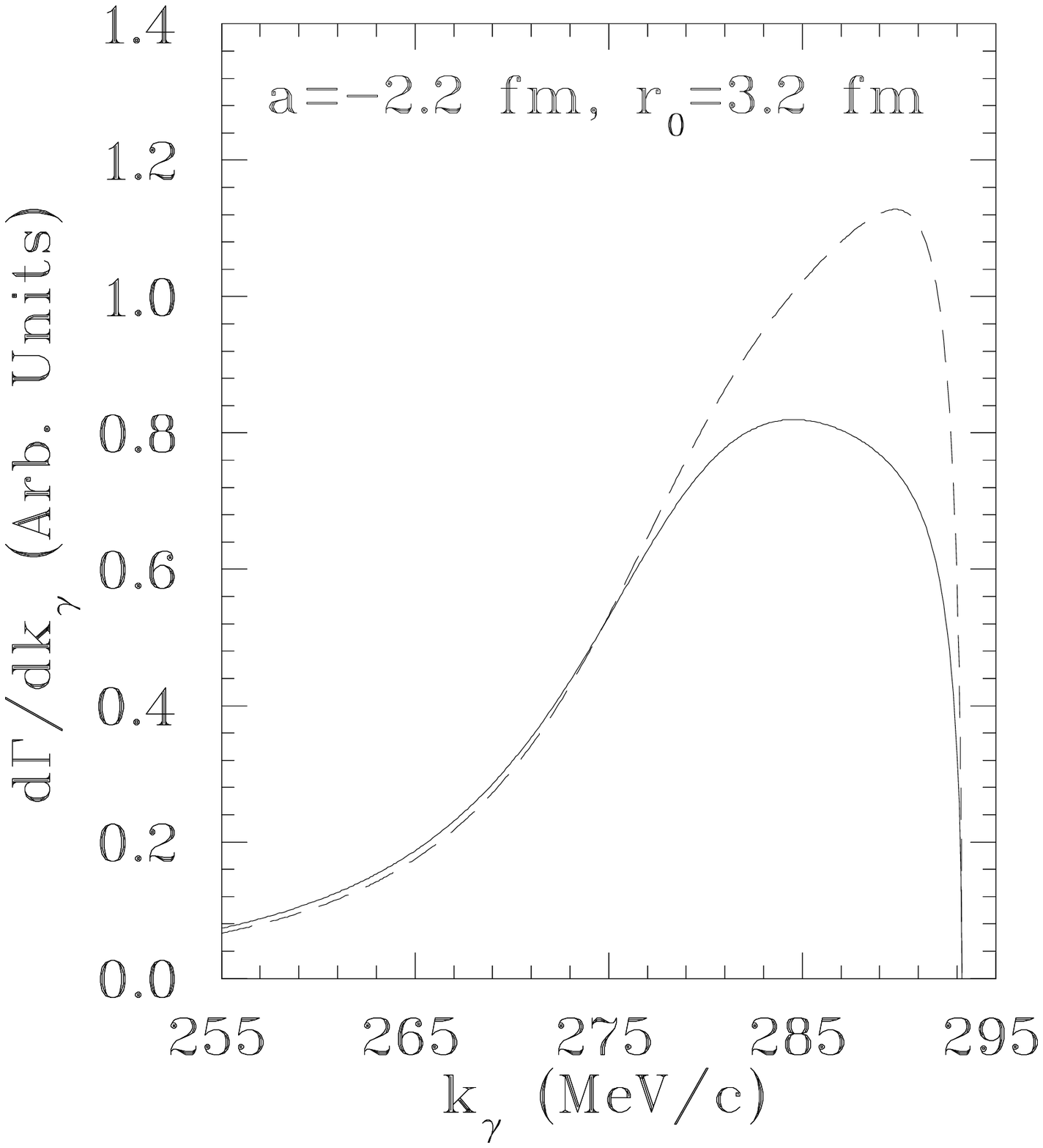}
\caption{Comparison of the spectrum from the $S_z=0$ state with
that from the non-zero initial projections. The dashed curve is the
spectrum from the $S_z=0$ magnetic sub-state of the deuteron.
\label{m01}}
\end{figure}

The S and D states of the deuteron interfere coherently,
with different coefficients depending on whether the initial state 
is $S_z=0$ or $S_z=\pm 1$. 
For this reason the shape of the spectrum is different for the 
transition from the $S_z=0$ sub-state than from the $S_z=\pm 1$ states
due to the D-state contribution.  Figure \ref{m01} shows
a typical result for the pure $\ell=0$ contribution calculated
with an asymptotic $\Lambda$-n wave function.

\section{Solutions with Exponential Potentials}\label{jost}

Rijken \ea\cite{rijken} have recently fit potential models to the 
$\Lambda$-nucleon scattering data.  We use the phase shifts
for the s wave determined by this group to define our potentials
so that the asymptotic form which is singular at the origin 
can be replaced with a more realistic wave function.

We will represent the effective interaction as a sum of exponential 
potentials. To this end we write the true potential as a Laplace 
transform

\eq
V(r)=\int_0^{\infty} \lambda(\mu)e^{-\mu r}d\mu.
\qe
Consider a calculation of this integral using a
Gauss-Laguerre integration scheme with a 5 point approximation.  
The integral has maximum precision for only a single value of $r$, 
and the value that is chosen as typical for $r$ represents the 
scale at which the integration is made.  We choose to work, 
primarily, at the scale of 1 fm, in which case the points in the 
integration scheme (normally dimensionless) represent inverse 
ranges in units of fm$^{-1}$.  The smallest of these inverse ranges 
has a value of 0.26356 fm$^{-1}$.  Since no exchange with such a small 
mass is believed to take place in this reaction, one expects to 
find zero for the coefficient of this term. Indeed, fits to the phase 
shifts produce very small numbers.  We set these values to zero and 
consider fits with only 4 parameters. Thus, the potential is 
parameterized with the form
\eq
V(r)=\sum_{i=1}^4\lambda_ie^{-\mu_i r}.
\qe
In this case the longest range entering into the problem is
that of the second Gauss-Laguerre point, which has an inverse range 
of 1.41340 fm$^{-1}$ and corresponds very well with two 
pion masses. The other ranges are 3.59642 fm$^{-1}$, 7.08581 fm$^{-1}$ 
and 12.64080 fm$^{-1}$.
We have also considered the scale of one-half fm (which means
that the inverse ranges are doubled in value) and 2 fm
(which means that the inverse ranges have been multiplied
by 1/2). 
While we work primarily with the fits on the scale of 1 fm,
the alternate fits provide estimates of the model dependence.  It is 
the 2 fm range fit which is more
useful for reasons discussed in section \ref{results}.

\subsection{Jost Solutions}

Consider the solution of the Schr\"odinger equation for a sum of exponential
potentials:

\begin{equation} 
V(r)=\sum_{j=1}^N \lambda_je^{-\mu_jr}.
\end{equation}
Jost \cite{jost} writes the solution for the s-wave, $f(p,r)$, as

\begin{equation} 
f(p,r)=e^{-ipr}\sum_{\gamma}C_{\gamma}(p)e^{-m_{\gamma}r} 
\end{equation}
where the subscript $\gamma$ is a compound index representing  a
set of N integers.  For example, for a three-term potential

\begin{equation} 
\gamma\equiv [j,k,l],\ \ j,k,l=0,1,2 \dots . 
\end{equation}
The coefficients $C_{\gamma}(p)$ are given by the recursion relation
\begin{equation} 
C_{[j,k,l]}(p)=\frac{\lambda_1C_{[j-1,k,l]}(p)+
\lambda_2C_{[j,k-1,l]}(p)
+\lambda_3C_{[j,k,l-1]}(p)}{m_{\gamma}(m_{\gamma}+2ip)},\label{recur} 
\end{equation}
where

\begin{equation} 
m_{\gamma}\equiv m_{[j,k,l]}\equiv j\mu_1+k\mu_2+l\mu_3 .
\end{equation}
The recursion is started with

\begin{equation} 
C_{[0,0,0]}=1, \ \ C_{[-1,k,l]}=C_{[j,-1,l]}=C_{[j,k,-1]}=0
\end{equation}
and is built up by first computing all coefficients with the sum of
indices equal to one, then two, {\it etc.} with no negative index.

The solution with the proper boundary condition at the origin 
for an incoming spherical wave with unit amplitude at 
infinity is

\begin{equation} 
\psi(p,r)=-\frac{f(p,r)-S(p)f(-p,r)}{2ipr} \label{psis}, 
\end{equation}
where
\begin{equation} 
S(p)=\frac{f(p,0)}{f(-p,0)}. 
\end{equation}
These expressions can be used to calculate the values of the S-matrix 
for any value of $p$.  

In order to calculate the overlap with the deuteron wave function, 
we require the wave function for real (positive) values of $p$.  
In this case we can write (for real $\lambda_j$)
\begin{equation} 
f(-p,r)=f^*(p,r)\ \ {\rm with}\ \ S(p)=e^{2i\delta(p)}.
\end{equation}
We can now write the wave function (Eq. 
\ref{psis}) as
\begin{equation} 
\psi(p,r)=-\frac{e^{i\delta(p)}}{2ipr}\left[
e^{-i\delta(p)}e^{-ikr}\sum_{\gamma}C_{\gamma}(p)
e^{-m_{\gamma}r}-
e^{i\delta(p)}e^{ikr}\sum_{\gamma}C_{\gamma}(-p)
e^{-m_{\gamma}r}\right] .
\end{equation}
Note that the lowest order term is given by
\eq
\frac{e^{i\delta(p)}\sin (pr+\delta(p) )}{pr},
\qe
which is identical to the asymptotic wave function.
We shall use these results to represent the $\Lambda$-n wave functions.

\subsection{Fits to the Phase Shifts}

\begin{figure}[p]
\epsfysize=180mm
\epsffile{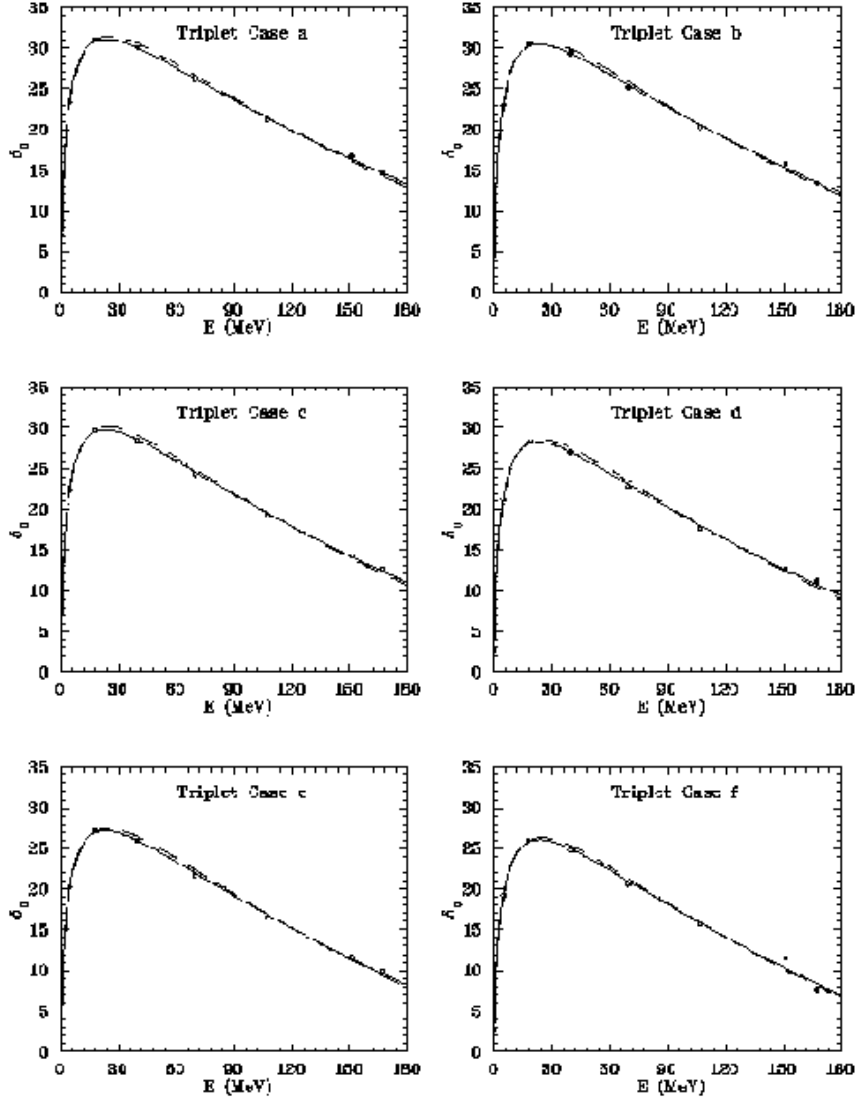}
\caption{Comparison of the fits for the s-wave phase 
shifts with the results of Ref. \protect\cite{rijken}
for the triplet states.  The results for the scale of
1 fm is represented by the solid line and the scale
of 2 fm by the dashed line.}
 \label{phase3}
\end{figure}

\begin{figure}[p]
\epsfysize=180mm
\epsffile{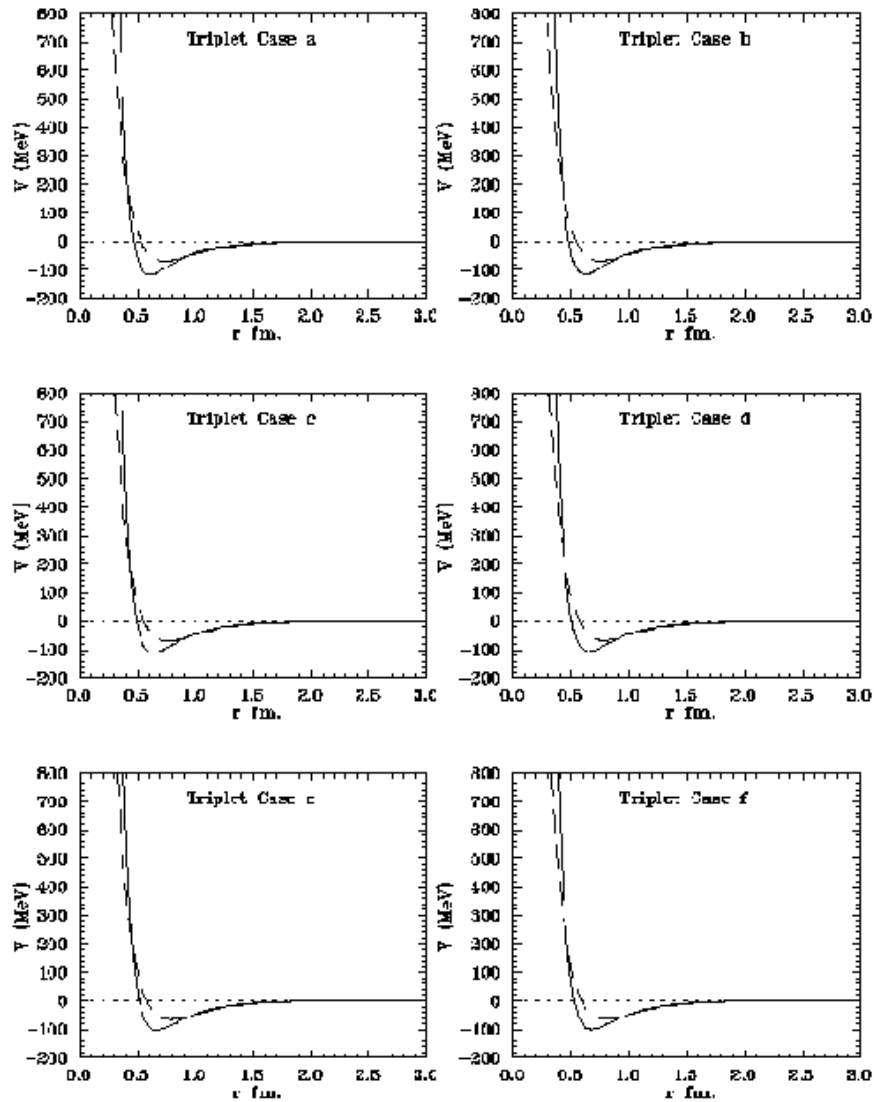}
\caption{Potentials obtained from the fits for the 
triplet case.  The meaning of the solid and
dashed lines is the same as in Fig. \protect\ref{phase3}.
\label{pot3}}
\end{figure}

Potentials were fit to the phase shifts\cite{stoks} for each of 
the six cases of Ref. \cite{rijken}, for both the singlet and triplet 
states and for each of the scales mentioned above.  Even
though the interaction of interest is for $\Lambda$-neutron
scattering, the fits were made to the $\Lambda$-proton phase
shifts since they were more closely related to data. The results
are shown in Fig.  \ref{phase3} for the triplet case.  The potentials
themselves are shown in Fig. \ref{pot3}.
Table \ref{potpars} lists the strengths from the fit for the scale
of 1 fm. It is this fit which is used in the remainder of the
paper unless otherwise noted.

\begin{table}\hspace{1.2in}
\begin{tabular}{ccccc}
\multicolumn{5}{l}{\bf \hspace{1.5in}  Singlet}\\
Case&$\lambda_1$&$\lambda_2$&$\lambda_3$&$\lambda_4$\\
a  &  --0.26431 &  --45.03399 &  304.25626 & 2327.48169\\
b  &  --0.25513 &  --52.52803 &  352.90689 & 3036.24731\\
c  &  --0.30427 &  --59.00891 &  396.75662 & 3540.90601\\
d  &  --0.26357 &  --72.61646  & 499.84998 & 5000.94873\\
e  &  --0.26643 &  --79.84088 &  561.22314 & 5784.05615\\
f  &  --0.27467 &  --87.65697 &  643.43768 & 6791.61670\\
\multicolumn{5}{l}{\bf \hspace{1.5in} Triplet}\\
a  &  --0.43381  & --35.94874  &  -0.17723 & 2502.01367\\
b  &  --0.42795  & --37.85583   &  9.45284 & 2803.92017\\
c  &  --0.43581  & --39.68054  &  32.58380 & 2944.07300\\
d  &  --0.44061  & --41.15301  &  52.91377 & 3138.80127\\
e  &  --0.40514  & --43.49734  &  77.60744 & 3344.70093\\
f  &  --0.35775  & --46.00097  & 104.79501 & 3541.92969\\
\end{tabular} 
\caption{Values of the parameters used in the fit for the
scale of 1 fm.} 
\label{potpars}\end{table}

Table \ref{ar} summarizes the scattering lengths and effective
ranges corresponding to these fits.  Also given are the
scattering lengths found by Rijken \ea with the original
potential.

\begin{figure}[p]
\epsfysize=180mm
\epsffile{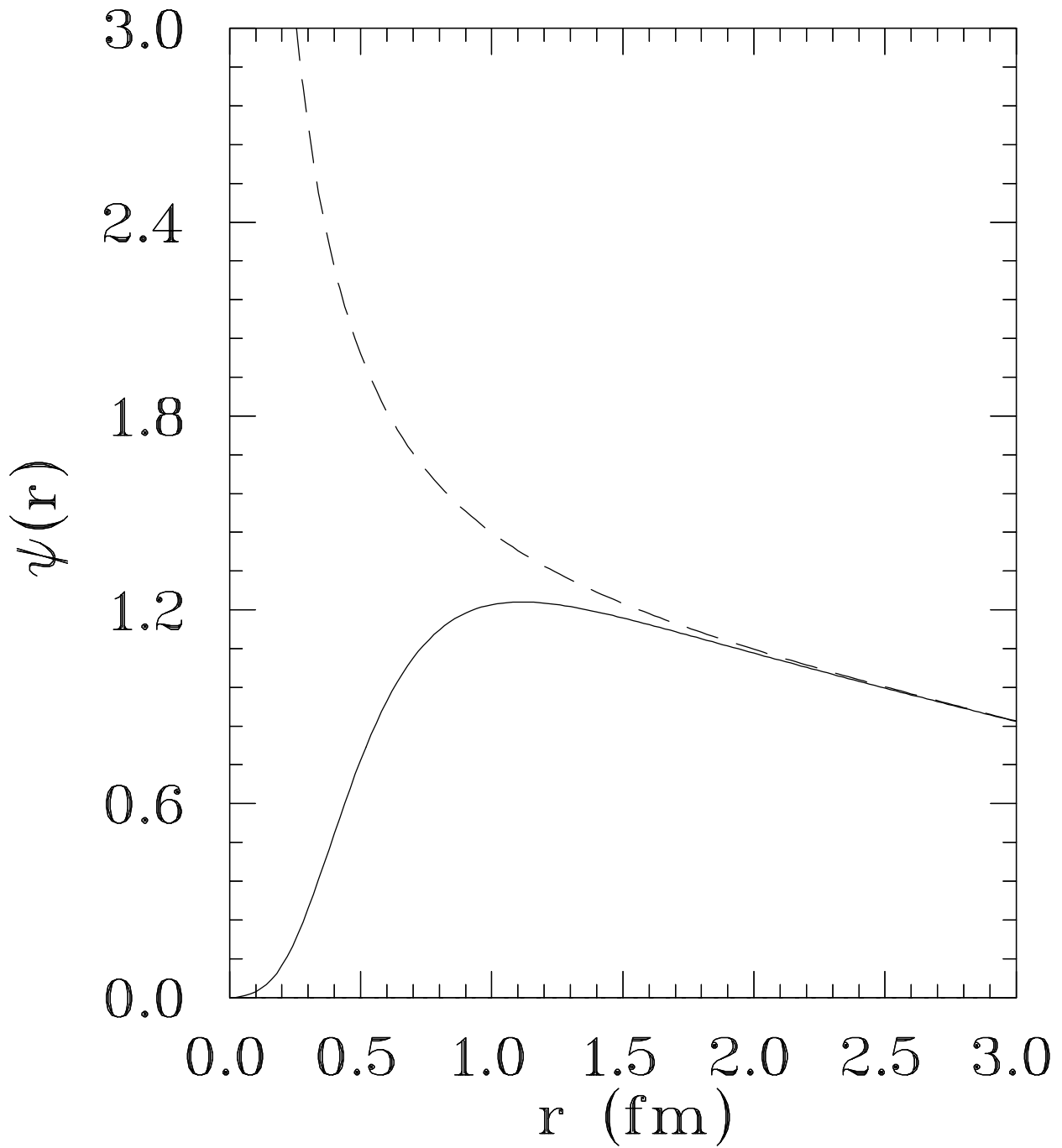}
\caption{Wave function for the asymptotic wave function
(dashed curve) compared with the Jost solution (solid curve). 
\label{wf}}
\end{figure}

\begin{table}\hspace{1.2in}
\begin{tabular}{cccccc}
\multicolumn{5}{l}{\bf \hspace{2.3in}  Singlet}\\
{\rm Case}&$a$ (Rijken \ea)&$a$ (1 fm)&$r_0$ (1 fm)
&$a$ (2 fm)&$r_0$ (2 fm)\\
a  & --0.71  & --0.73  & 6.71& --0.63 & 4.58 \\
b  & --0.90  & --0.92  & 5.57& --0.82 & 3.99 \\
c  & --1.20  & --1.23  & 4.55& --1.11 & 3.52 \\
d  & --1.71  & --1.74  & 3.71& --1.59 & 2.99 \\
e  & --2.10  & --2.16  & 3.46& --1.97 & 2.57 \\
f  & --2.51  & --2.59  & 3.29& --2.36 & 2.68 \\
\multicolumn{5}{l}{\bf \hspace{2.3in} Triplet}\\
a  &  --2.18 & --2.21  & 2.99& --2.07 & 2.57 \\
b  &  --2.13 & --2.16  & 3.05& --2.03 & 2.71 \\
c  &  --2.08 & --2.10  & 3.15& --1.97 & 2.68 \\
d  &  --1.95 & --1.97  & 3.33& --1.84 & 2.84 \\
e  &  --1.86 & --1.86  & 3.40& --1.74 & 2.94 \\
f  &  --1.75 & --1.74  & 3.42& --1.65 & 3.03 \\
\end{tabular} 
\caption{Values of the scattering length ($a$) and 
effective range ($r_0$) from the fits described.} 
\label{ar}\end{table}
In Fig. \ref{wf} are plotted the full wave function and the
asymptotic wave function for the singlet ``a'' case of Ref. 
\cite{rijken}.

\section{Results\label{results}}

\begin{figure}[p]\epsfysize=180mm\epsffile{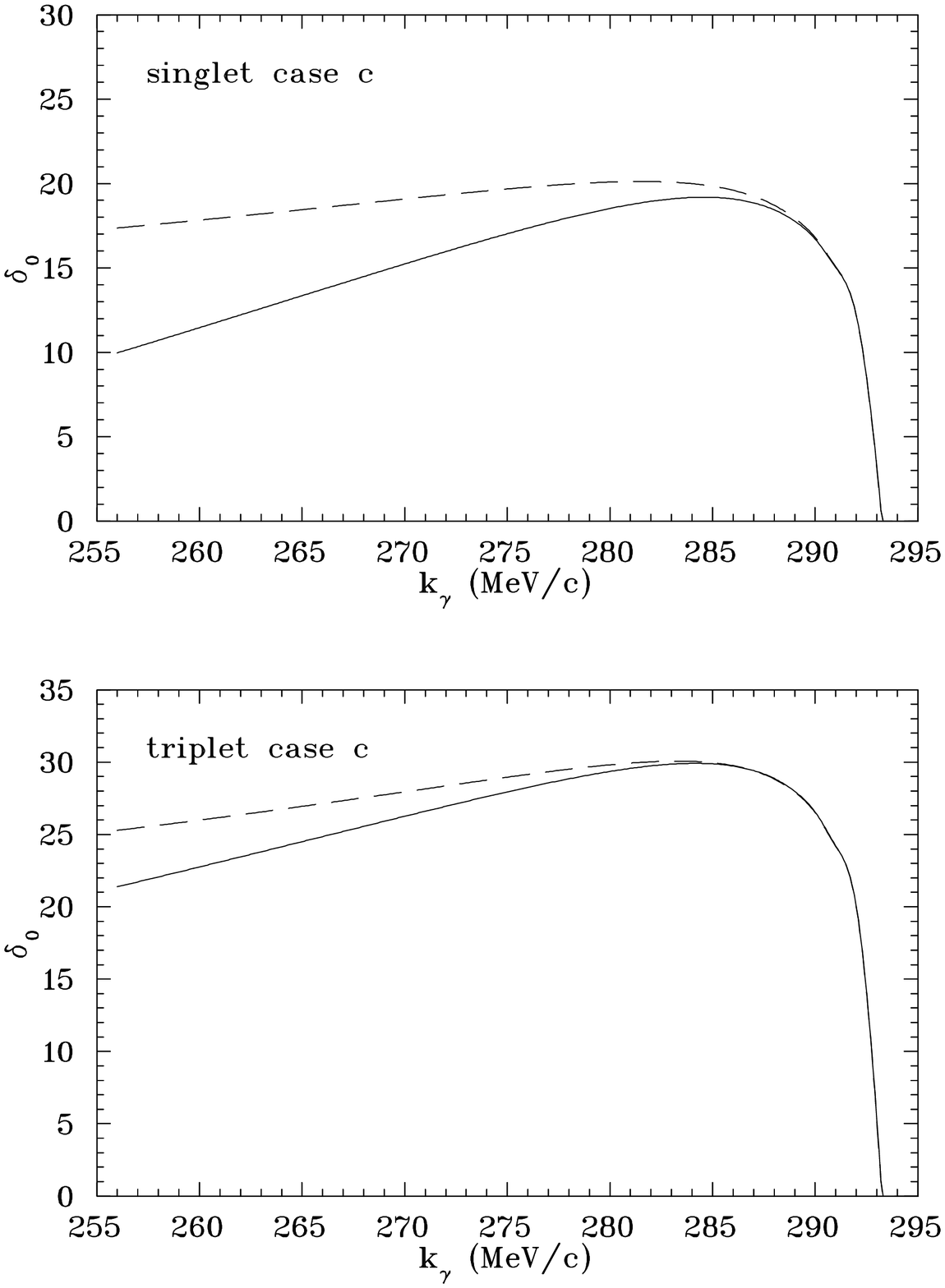}
\caption{Phase shifts from the fitted potentials (solid) and from the
effective range expansion (dashed) for the case c of Rijken \ea\ 
for the singlet and triplet cases. \label{phase}}
\end{figure}

We now turn to problems in the analysis 
and some possible solutions.  The first
issue is the use of the effective range expansion (ERE).
In the $\pi^-d\rightarrow nn\gamma$  measurement the ERE is 
adequate to describe the
s-wave phase shift over the full range of data being analyzed. As is 
shown in Figure \ref{phase}, this may not be the case
for the n$\Lambda$ measurement. Certainly it is not adequate 
over the full range, and it may be somewhat questionable even if only
the upper 10 MeV of the spectrum is used. Of course the ERE is
only useful if the asymptotic wave functions are used or if the
short-range corrections are made with a technique similar to
that employed in Ref. \cite{ggs}.

\begin{figure}[p]
\epsfysize=180mm
\epsffile{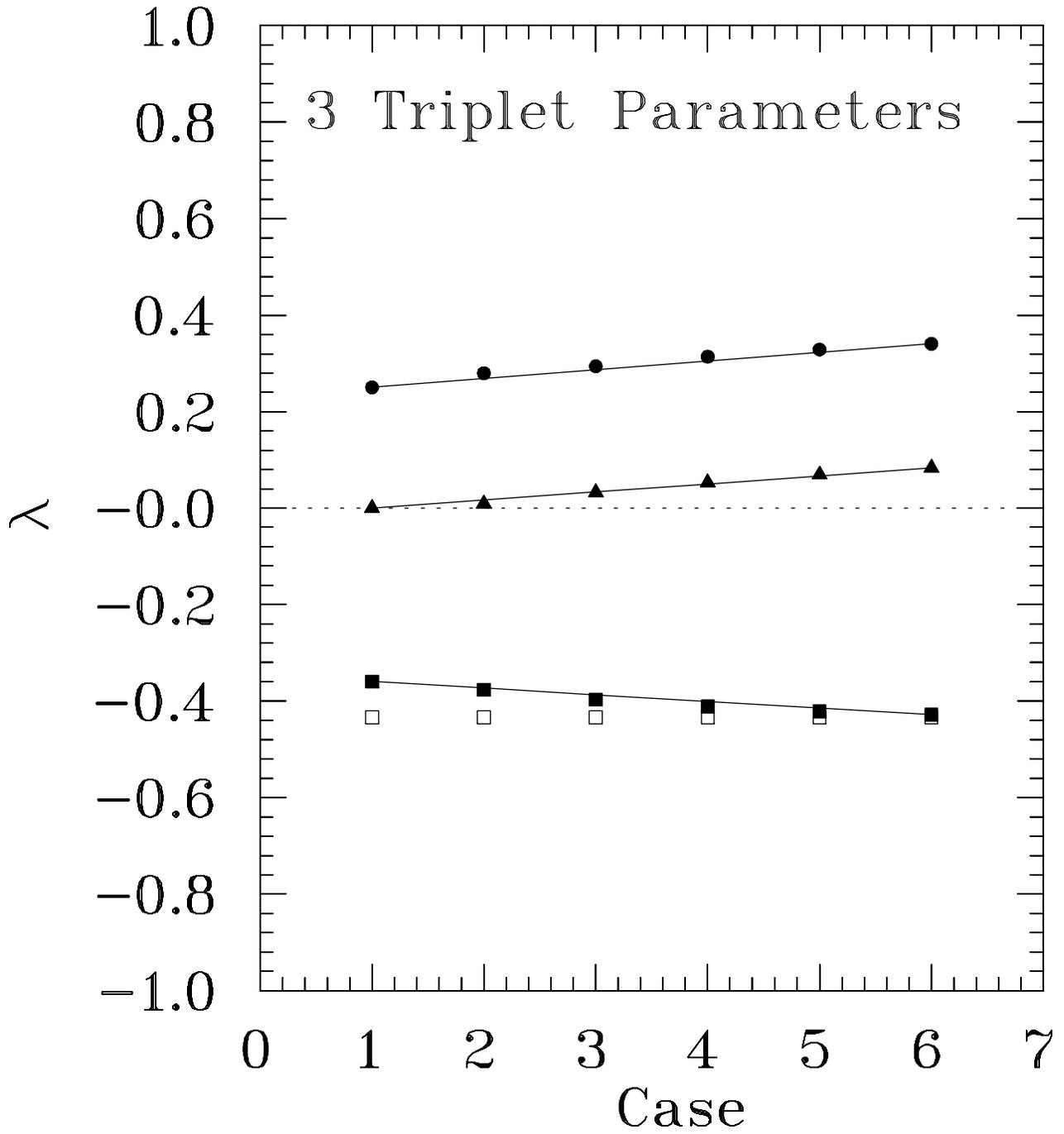}
\caption{Parameters from the fit to the phases shifts of Rijken et
al. showing the linear representation for the triplet case. 
\label{pars3t}}
\end{figure}

To find an alternative to the ERE we examined the parameters
which came from the fitting of the exponential potentials
to the phase shifts of Rijken \ea\  It was found that
an acceptable fit could be obtained with the longest range parameter
held fixed.  Plots of the other three parameters (see Fig.  
\ref{pars3t} for the triplet case) show a linear behavior with case
number.  We therefore defined continuous variables, which
reproduce the phase shifts (to a good approximation)
at the integers 1 through 6 (corresponding to cases a through f), 
and create an interpolation and extrapolation procedure.

We define strengths
\eq 
\lambda_i=c_i+x d_i 
\qe
for i=2, 3, 4.  Here $x=x_s$ for the singlet case and $x=x_t$ for
the triplet case.  The values of $c_i$, $d_i$ and $\lambda_1$ are given
in Table \ref{cds}.  While this method allows a range of scattering
lengths to be produced with an appropriate short-range wave
function, there are limitations.  Since we would like to be able
to study scattering lengths outside the range given in Ref. 
\cite{rijken}, we wish to consider values of $x$ outside the range
(1-6).  Indeed this does give an extension of the range of
scattering lengths, but for the 1 fm scale the range is limited.
Outside this range the scattering lengths return to previous values.
This constraint dictates the limits of the analysis shown in the
figures to follow.  It was found for the fits on the scale 
of 1/2 fm that the problem was greater, so that the analysis would be 
restricted to an even smaller area.  The 2 fm range
fits were used for the model dependence estimates because, in
this case, a larger range of scattering lengths was feasible.
However, the scattering lengths do not reproduce those of
Ref. \cite{rijken} as well as the 1 fm scale, as may be seen in 
Table \ref{ar}. 

\begin{table}\hspace{1.2in}
\begin{tabular}{cccccccc}
&$\lambda_1$&$c_2$&$d_2$&$c_3$&$d_3$&$c_4$&$d_4$\\
Singlet&--0.26431&--36.410&--8.624&235.36&68.90&1457.6 & 869.9\\
Triplet&--0.43381& --34.576&--1.373&--16.90&16.73&2321.4& 180.6\\
\end{tabular} 
\caption{Values of the parameters for the linear representation
of the potentials.} \label{cds}\end{table}

\begin{table}[htb]\hspace{1.2in}
\begin{tabular}{cccccc}
\multicolumn{6}{l}{\bf \hspace{1.5in}  Singlet}\\
{\rm Case}&$\lambda_1$&$\lambda_2$&$\lambda_3$&$\lambda_4$\\
1  &  --0.26431 &  --45.03399  & 304.25626 & 2327.48169\\
2  &  --0.26431  & --52.14749  & 349.97018 & 3001.88623\\
3  &  --0.26431 &  --61.00436 &  417.78851 & 3645.56372\\
4  &  --0.26431  & --72.61623  & 499.85406 & 5005.63232\\
5  &  --0.26431 &  --79.96191 &  562.18121 & 5807.02148\\
6  &  --0.26431  & --88.15266  & 648.76056 & 6839.37939\\
\multicolumn{6}{l}{\bf \hspace{1.5in} Triplet}\\
1  &  --0.43381 &  --35.94874 &   --0.17723 & 2502.01367\\
2  &  --0.43381  & --37.67183   & \ 9.01196  & 2795.74268\\
3  &  --0.43381  & --39.68257  &  32.59604 & 2939.50488\\
4  &  --0.43381  & --41.09288  &  52.92378 & 3138.51001\\
5  &  --0.43381 &  --42.13403 &   69.27229 & 3286.15967\\
6  &  --0.43381  & --42.81454   & 83.45535 & 3404.96362\\
\end{tabular} 
\caption{Values of the parameters obtained for the 1 fm fit
with constant $\lambda_1$.} \label{prampot}\end{table}

\begin{figure}[p]
\epsfysize=170mm
\epsffile{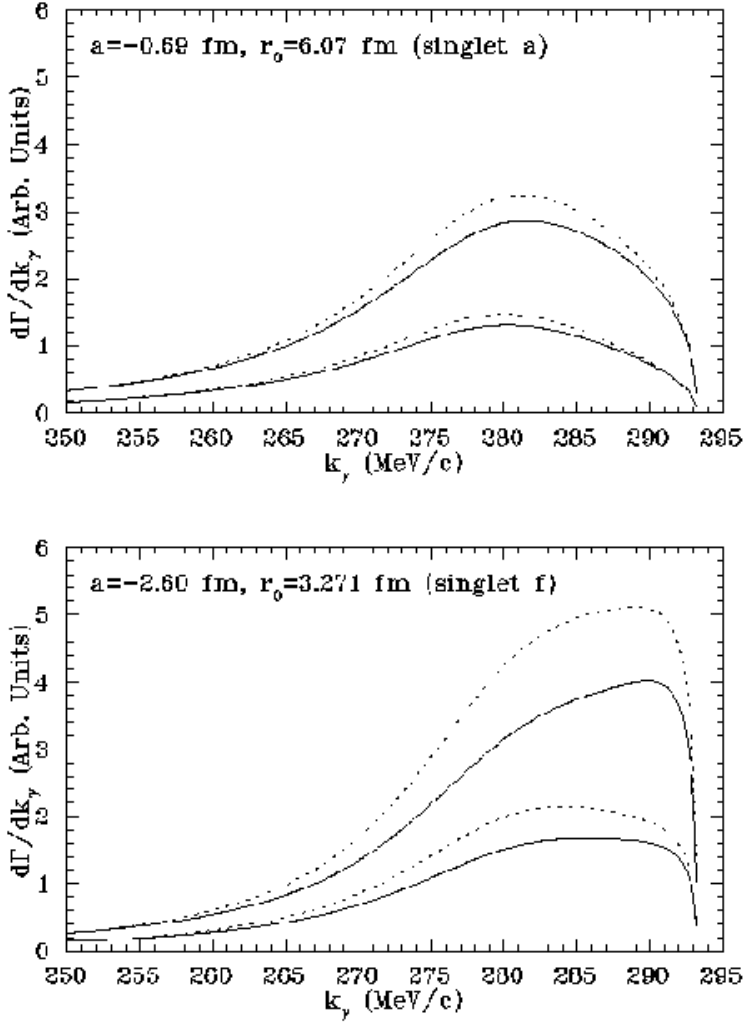}
\caption{Comparison of the spectra expected ($\ell=0$
contribution only) with an asymptotic wave function 
(solid curves) with that using a solution to the Schr\"odinger 
equation (dotted curves) for the Jost parameters corresponding
to cases ``a'' and ``f'' for the singlet final state. The lower 
curves are those resulting for the singlet spectrum and the 
upper ones are for the triplet spectrum. 
\label{nlamjboth}}
\end{figure}

In performing the analysis one may use the
asymptotic wave functions or the full wave functions.  The 
use of the full wave functions should be preferable,
but one can obtain an idea of the sensitivity to the details
of the full wave function by comparing results with those for
the asymptotic wave functions.  
Figure \ref{nlamjboth} shows the sum of the $S_z=0$ and 
$S_z=\pm 1$ s-wave contributions to the spectra as shown in
Fig. \ref{m01} for the asymptotic wave functions (solid lines) 
and the full wave functions (dashed lines) for the parameters
of the singlet ``a'' and ``f'' cases. 

\begin{figure}[p]
\epsfysize=180mm
\epsffile{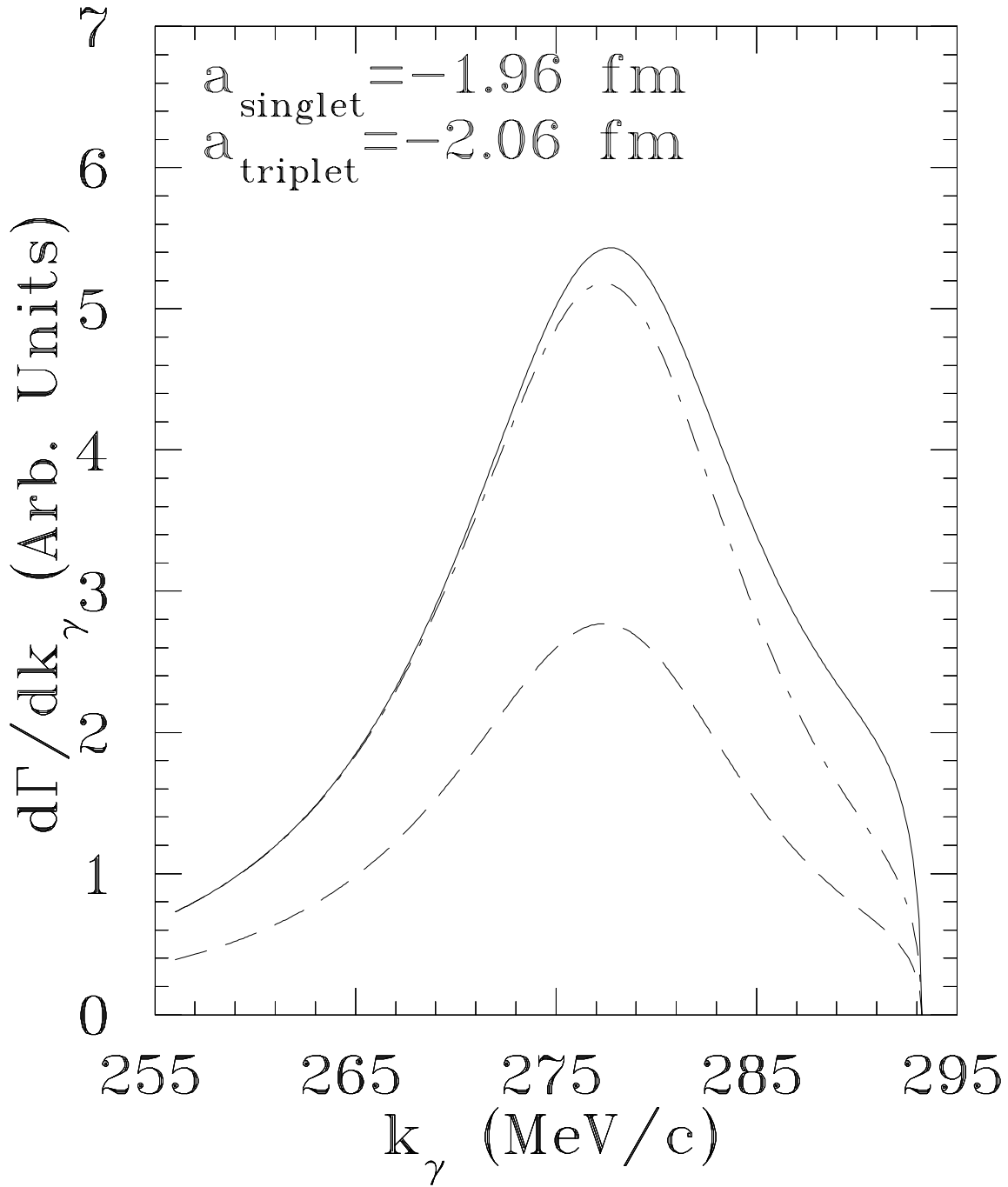}
\caption{Comparison of singlet (dashed) and triplet (solid) 
spectra for a full calculation with realistic wave functions.  
The dash-dot curve shows the singlet spectrum renormalized
to the triplet at the lowest energy.
\label{spect}}
\end{figure}

Perhaps the largest problem in the measurement is the separation
of the  singlet and triplet scattering lengths in the final state.  
If no spin degrees of freedom are measured then the singlet and 
triplet states contribute incoherently to the rate, and it might 
seem impossible to separate them.  However, because the interference 
between the S and D states of the deuteron differs, the shapes of 
the singlet and triplet spectra are different, as illustrated in Fig. 
\ref{spect}. Because the triplet gives a larger contribution than 
the singlet, a greater sensitivity to the triplet scattering is seen 
in the fits of the data and pseudo-data in the figures which follow.  

\begin{figure}[p]
\epsfysize=170mm 
\epsffile{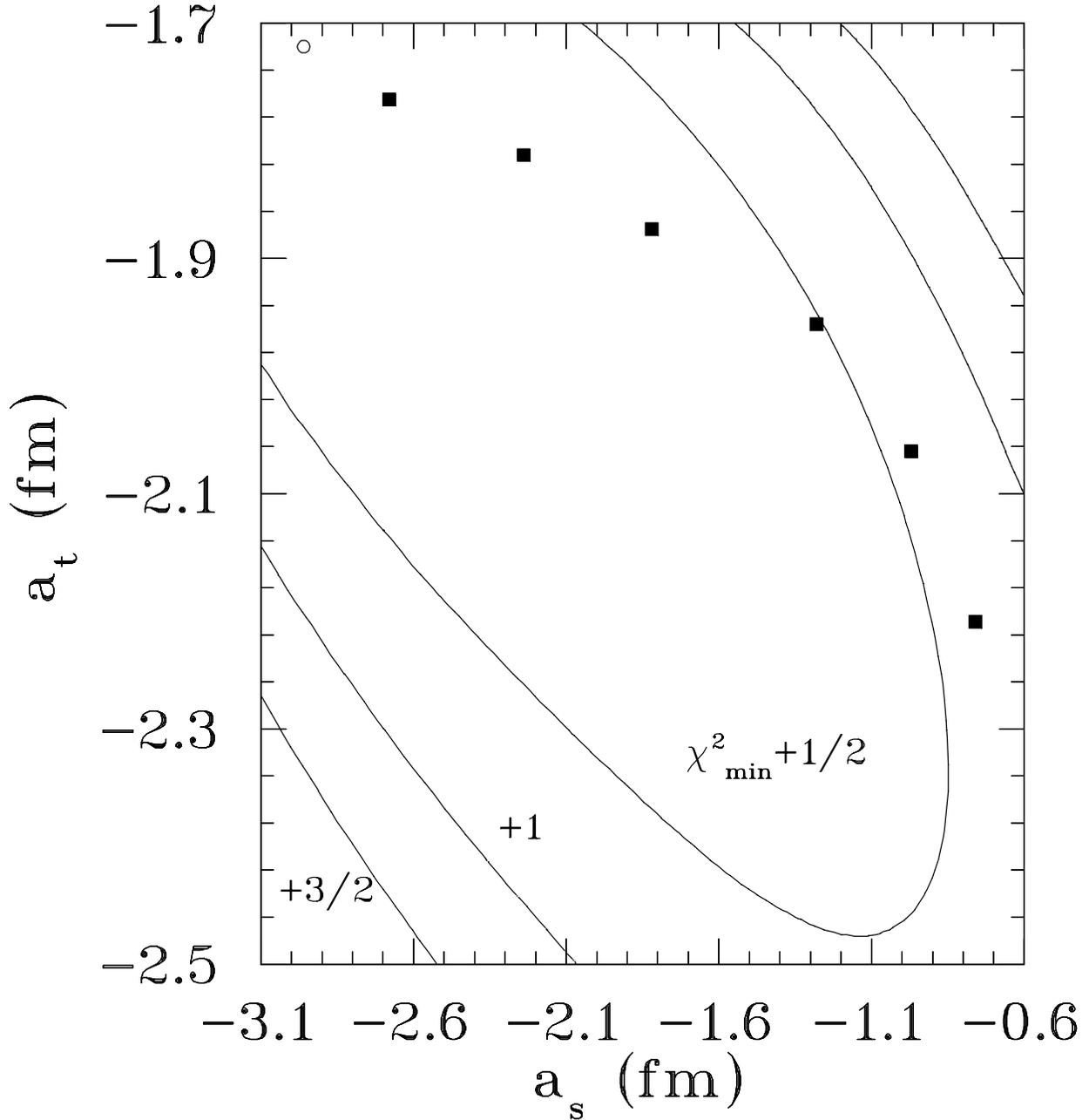}
\caption{Analysis of the data of Gall \ea\ using the 1 
fm scale fits.  The contours are lines of constant $\chi^2$.
Points lying inside the curve labeled +1 are within one standard
deviation of the minimum. The solid squares are the values taken
for the 6 cases of Ref. \cite{rijken} and the open circle is
the value at the minimum $\chi^2$ of the fit.
\label{conexpp}}
\end{figure}

Figure \ref{conexpp} shows an analysis of the data of Gall \ea 
\cite{gall}.  Since this
experiment was only a feasibility study, one can not expect to
obtain much information about the scattering lengths.  It is
interesting to see, however, that the values are in the range
expected. The open circle in the upper left hand corner 
shows the point of minimum $\chi^2$.

\begin{figure}[p]\epsfysize=180mm \epsffile{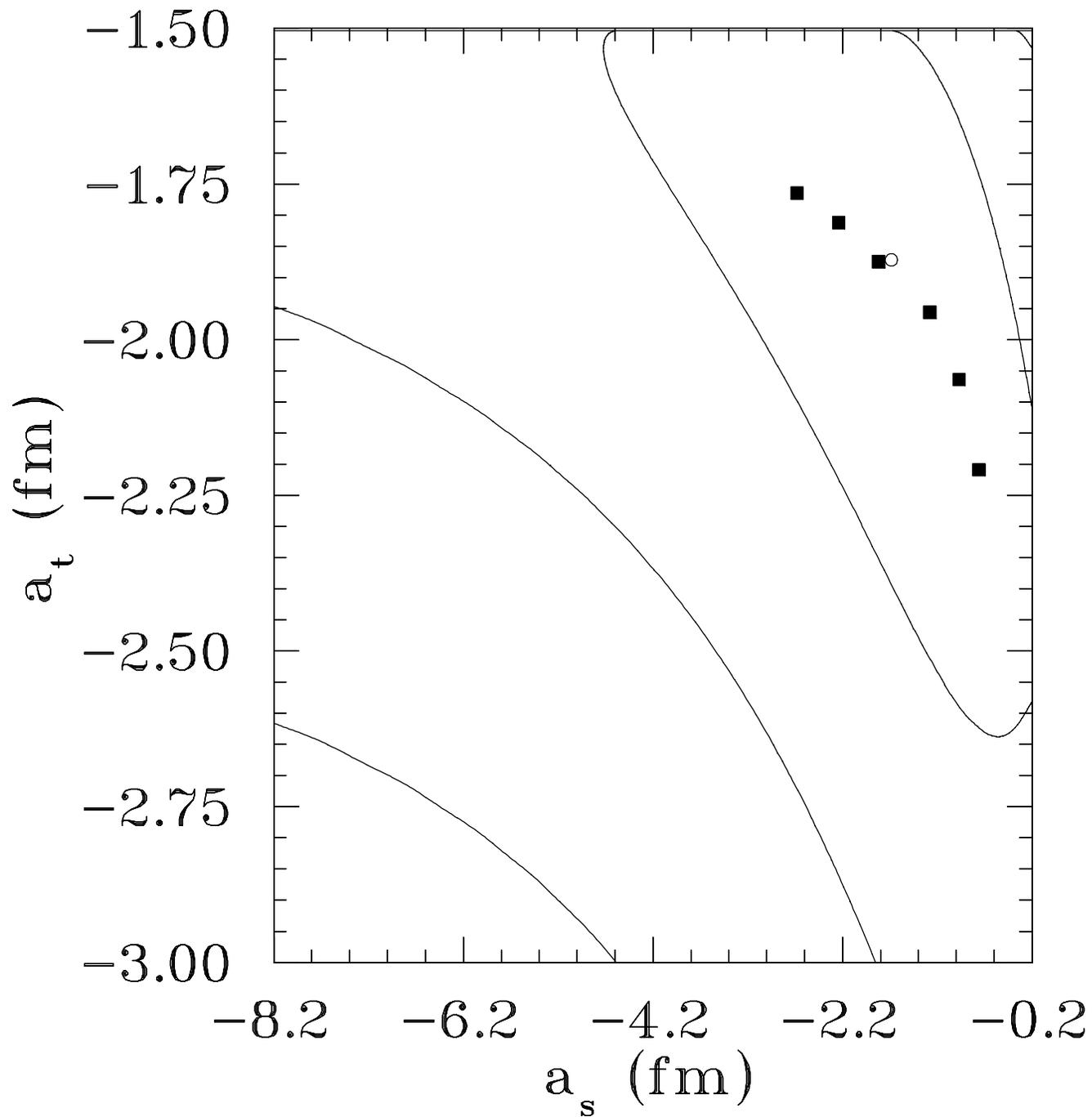}
\caption{Analysis of the data of Gall \ea\ with the
2 fm range fits. The innermost curve corresponds to one
standard deviation, the next to two, etc.
\label{conexplr}}
\end{figure}

Figure \ref{conexplr} shows a similar analysis of the Gall 
\ea\ data with the 2 fm scale fits. In this case a larger range 
of scattering lengths can be studied so that an error estimate 
can be given. One-standard-deviations limits can 
be read from the graph (the inner contour) with the triplet 
length lying between --1.3 and --2.6 fm and the
singlet length between --0.2 and --6.3 fm.

\begin{figure}[p]\epsfysize=180mm \epsffile{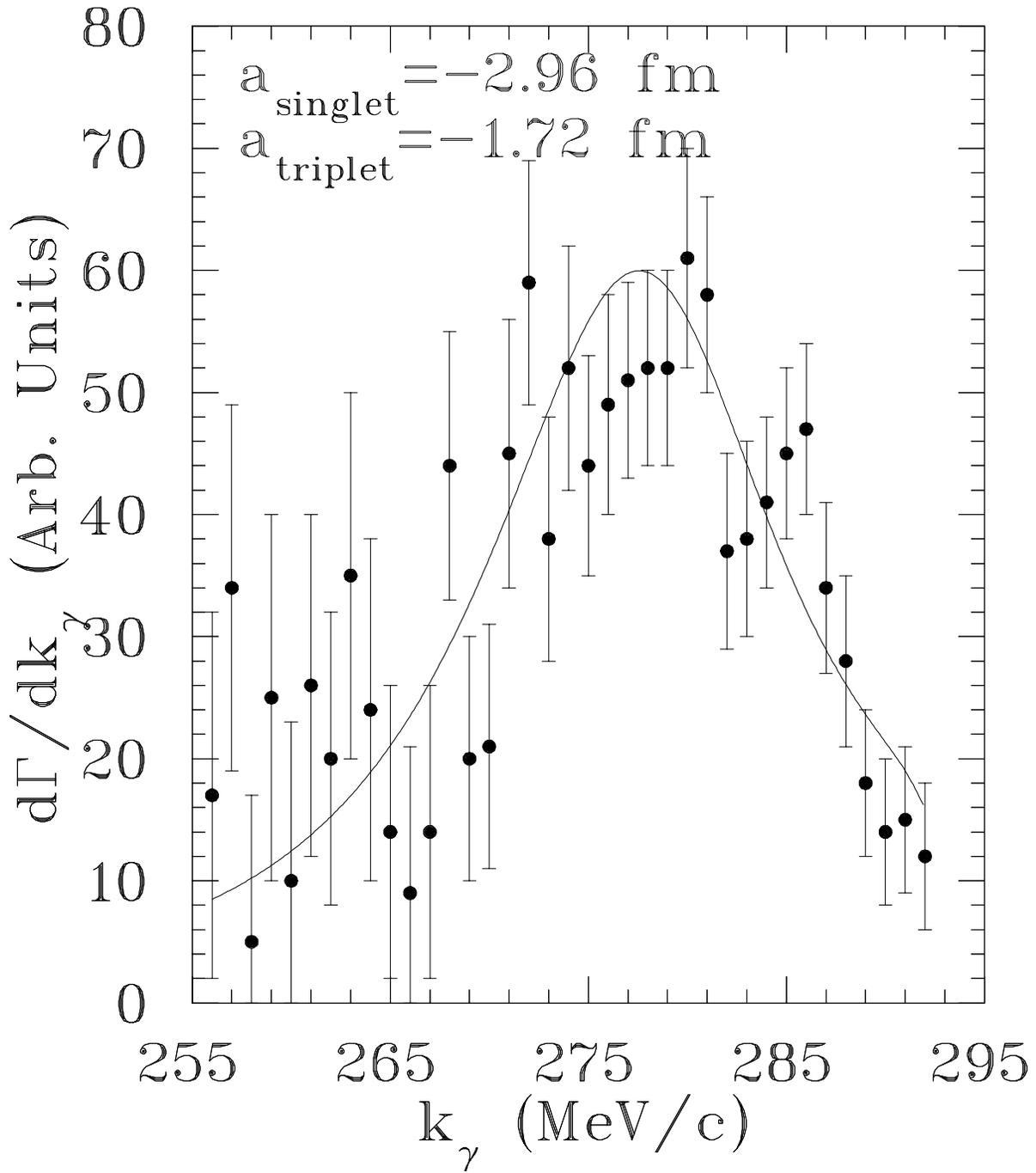}
\caption{Comparison with the data of Gall \ea\ with the
best fit using the 1 fm scale.
\label{cmpdata}}
\end{figure}

Figure \ref{cmpdata} provides a comparison of the experimental and 
theoretical spectra for the scattering lengths corresponding to 
the minimum $\chi^2$ (=37.8 for 37 data points) at $a_s=-2.96$ 
and $a_t=-1.72$ for the 1 fm scale.

\begin{figure}[p]
\epsfysize=170mm  
\epsffile{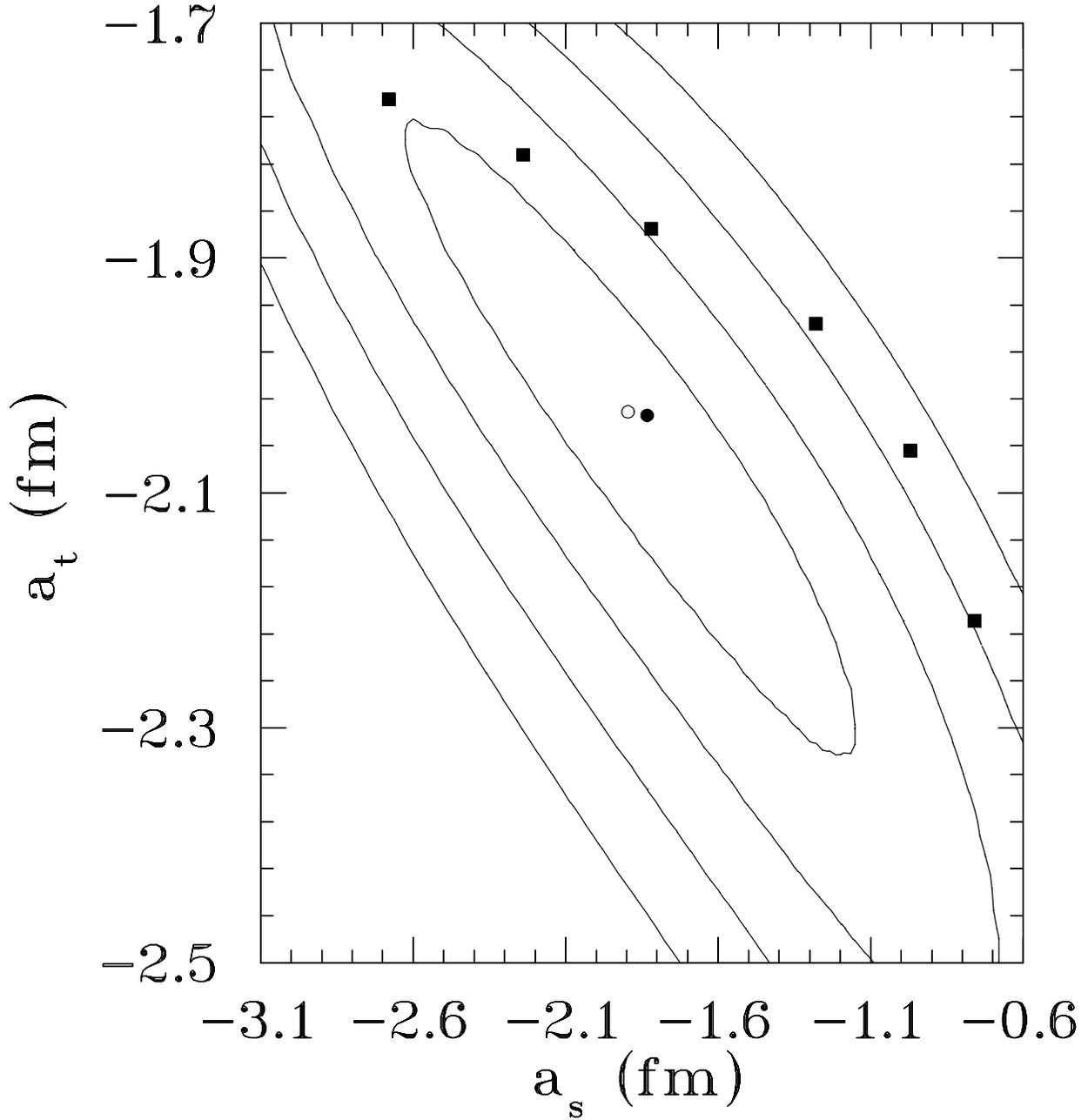}
\caption{Analysis with a full Jost model of a pseudo data set 
generated with 3\% errors. The 1 fm scale fit was used to 
generate the data and make the analysis. The solid circle is
the input value chosen for the test and the open circle is the
central value corresponding to the minimum $\chi^2$.
 \label{conexptstp}}
\end{figure}

Figure \ref{conexptstp} presents a similar analysis of a pseudo 
data spectrum. The pseudo data were generated from a selected 
spectrum by including errors chosen from a Gaussian 
distribution such that they have a value of 3\%
of the rate at the maximum and are proportional to
the square root of the rate at other points, as would
be the case with errors dominated by counting statistics.
The analyses shown were made over the full range 
of the Gall \ea\ data from 255 to 293 MeV unless otherwise
specified. The case chosen for presentation in Fig. 
\ref{conexptstp} is a favorable one in the sense that the
scattering lengths used for generation of the pseudo-data (solid
circle) fall near the center of the inner ellipse.  As is
necessary statistically, most of the results do fall nearer
the edge of this ellipse with about 1/3 falling outside. 
One can observe from this figure that the experimental 
uncertainty in such an experiment would be $\sim\pm 0.3$ fm 
for the triplet scattering length and $\sim\pm 0.8$ fm for 
the singlet. 

\begin{figure}[p]
\epsfysize=180mm  
\epsffile{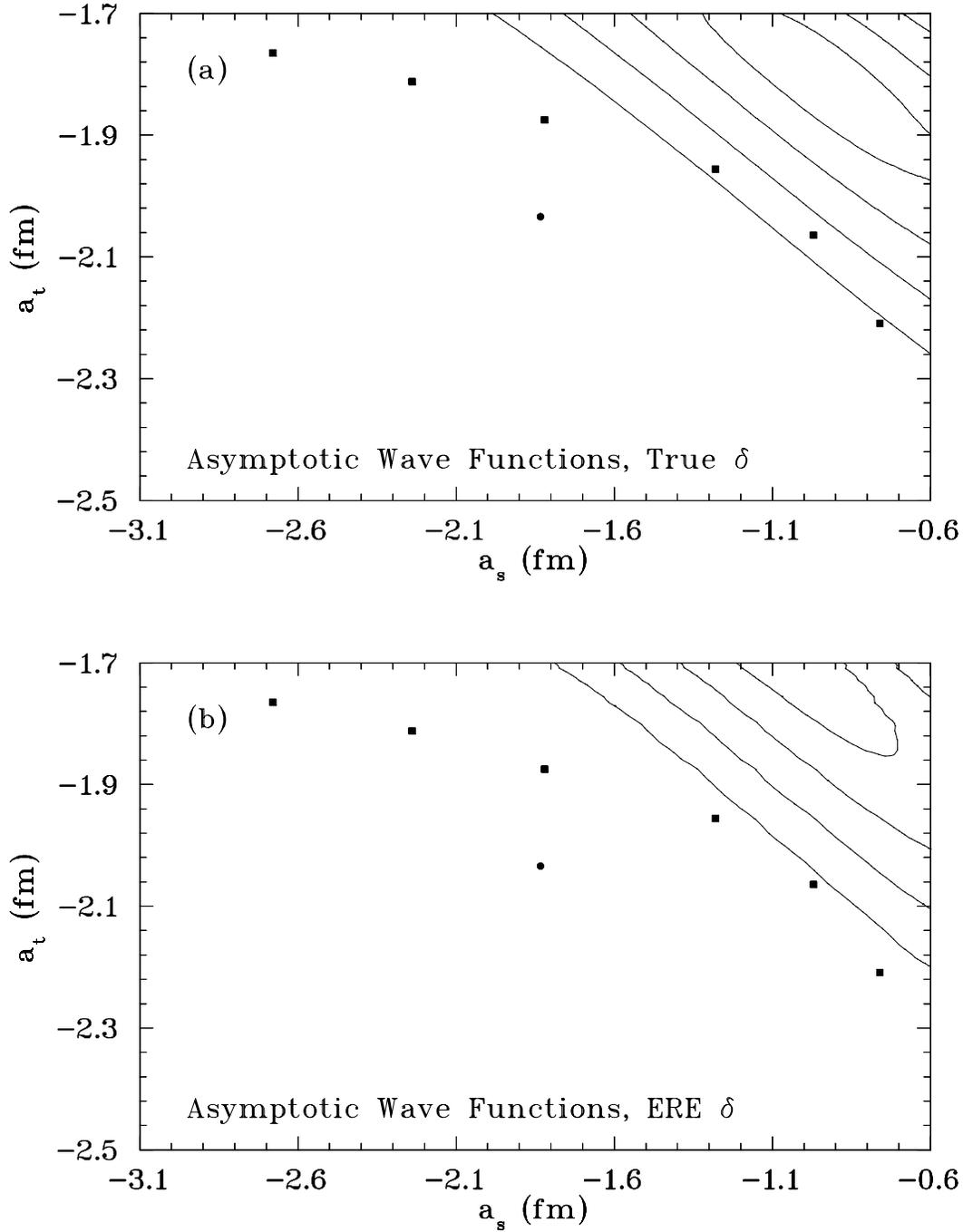}
\caption{Analysis of the same pseudo data set as in Figure
\protect\ref{conexptstp} with asymptotic wave functions.  In
the upper part the correct phase shifts are used while in
the lower portion the effective range expansion is used.
\label{contstasy}}
\end{figure}

It is possible to use such an analysis of pseudo data to estimate
errors due to model assumptions.  For the following we use
the same pseudo data that were described above.  Figure 
\ref{contstasy}a
shows the results of an analysis using asymptotic wave functions
with phase shifts determined from the Jost solutions. 
Figure \ref{contstasy}b shows an analysis with the additional 
approximation that the phase shifts are taken from the ERE.
As can be seen a significant error results.

\begin{figure}[p]
\epsfysize=180mm  
\epsffile{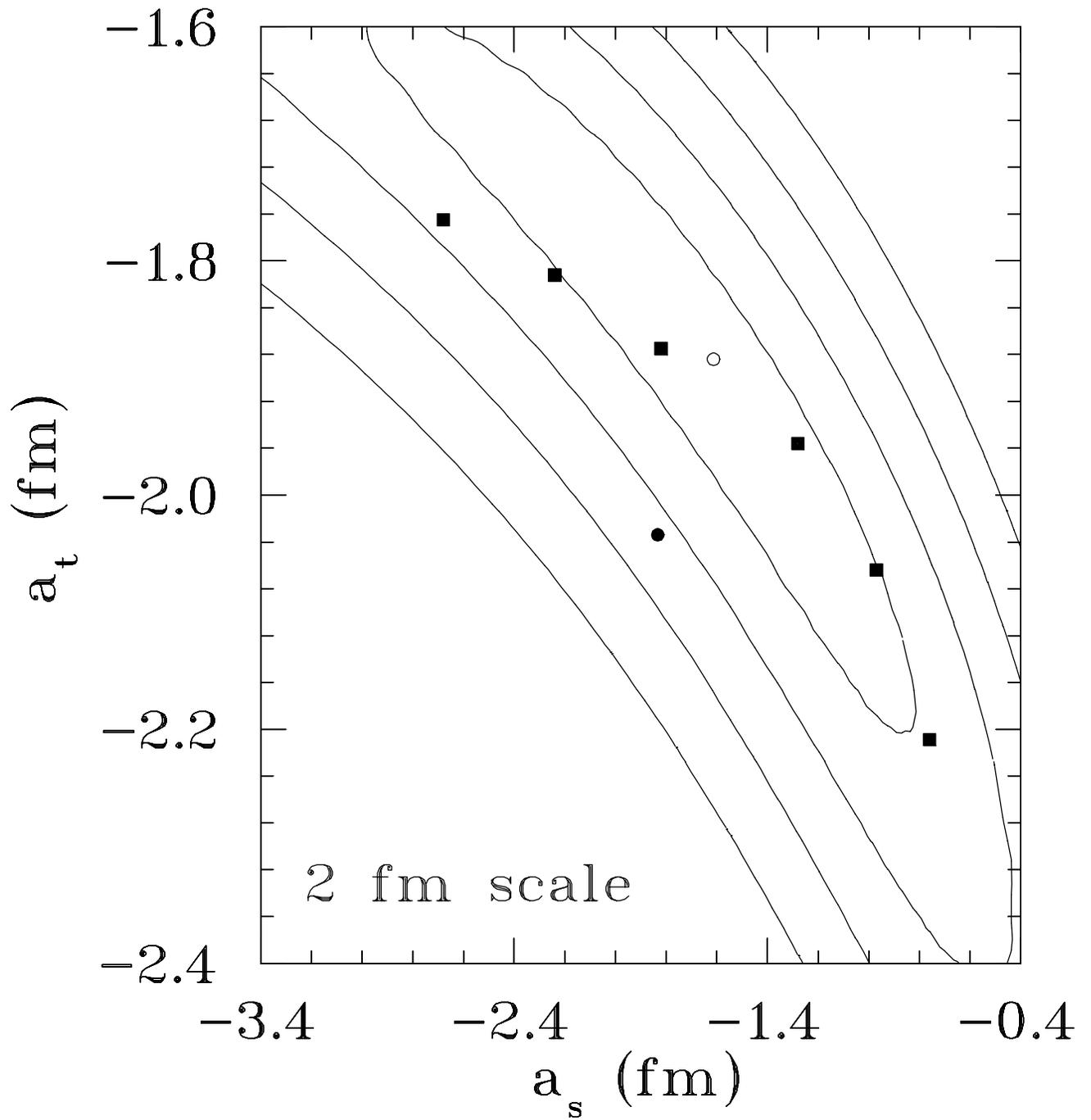}
\caption{Analysis of the same pseudo data set as in Figure
\protect\ref{conexptstp} with wave functions generated 
using the 2 fm range fit. \label{contstlr2}}
\end{figure}

Use of the asymptotic wave functions is an extreme approximation.
Although we do not know exactly which more realistic wave function
should be used, we can choose one among the Jost models we have 
been considering. To this end we analyze the spectrum created with 
the 1 fm scale using the 2 fm scale spectra.  The result can be
seen in
Fig. \ref{contstlr2}.  The positions of the singlet and triplet
scattering lengths at the minimum are --1.613 and --1.884 fm to
be compared with the values found with the (1 fm scale) analysis 
above of --1.896 and --2.031 fm (the input values are --1.968 and 
--1.827 fm). Thus we find model errors of 
$\pm0.28$ fm for the singlet and $\pm0.15$ fm for the triplet case. 
We will take these as estimates of the model (theoretical) errors.

\begin{figure}[p]
\epsfysize=180mm  
\epsffile{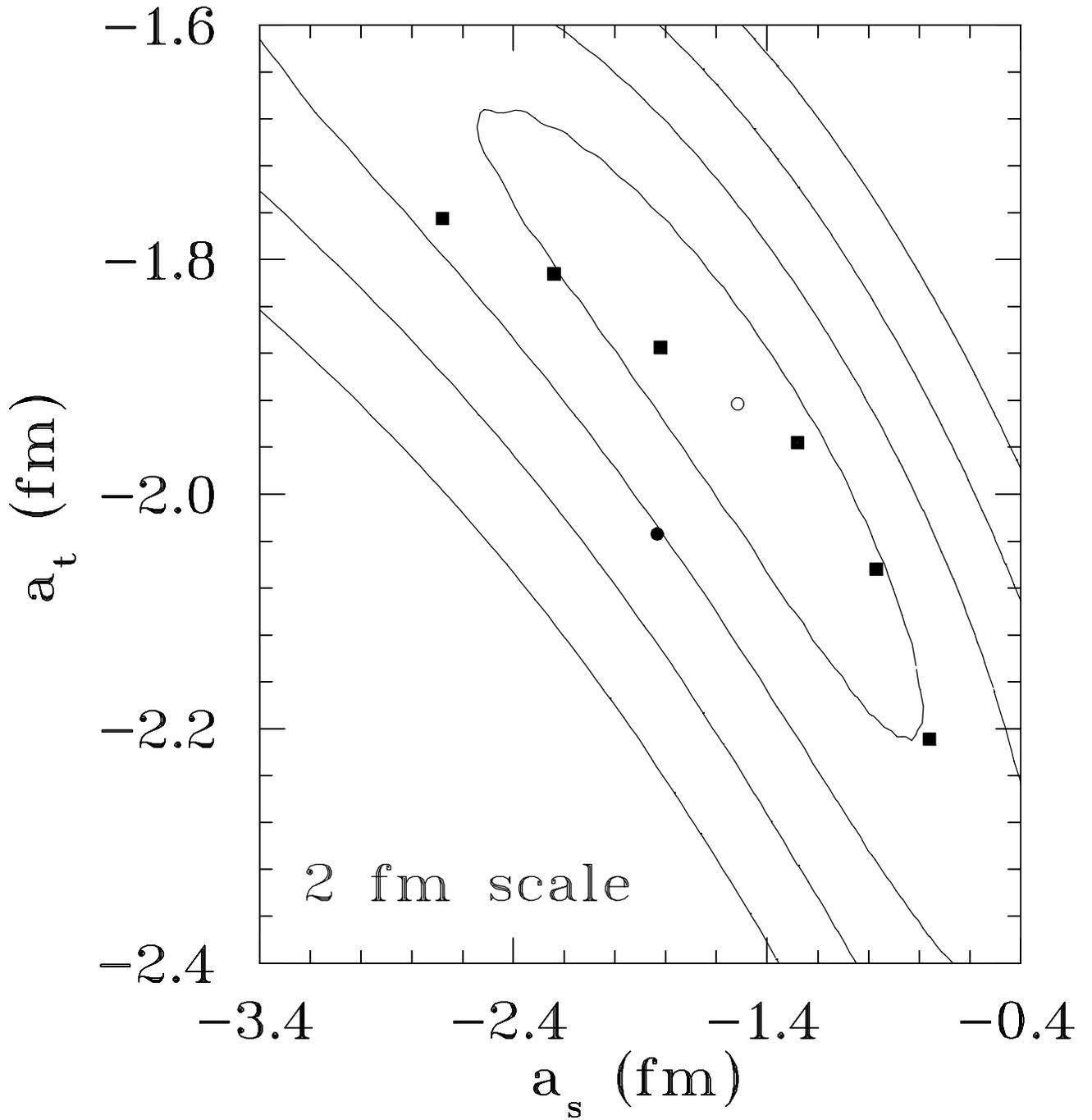}
\caption{Analysis of the same pseudo data set as in Figure
\protect\ref{conexptstp} with wave functions using the 
2 fm range fit over only the upper 10 MeV of the
spectrum. \label{contstlrup}}
\end{figure}

Figure \ref{contstlrup} shows the same type of analysis but
including only the upper 10 MeV of the spectrum in the 
comparison. It is seen that the results are very similar, 
due perhaps to the fact that the main difference in shapes
between singlet and triplet occurs in this region (see
Fig. \ref{spect}).  The analysis over this range may offer
experimental advantages in addition, since the background due
to neutral pion decay is less severe.

\begin{figure}[p]
\epsfysize=160mm
\epsffile{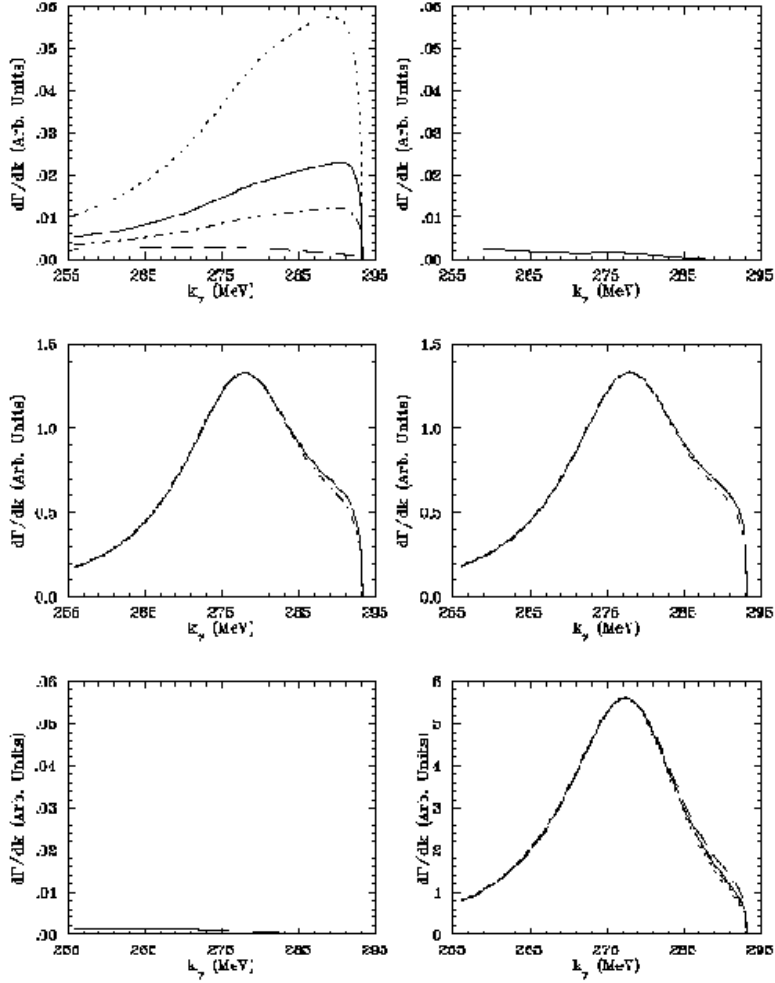}
\caption{Spectra for the various spin combinations of the
initial deuteron and the final photon.  The three figures
to the left are for right circular polarization and 
those to the left for left circular polarization of the final
photon.  The three rows are for $S_z=+1,\ 0$ and
$-1$ from the top down. The solid curve has $a_s=-1.15,\ 
a_t=-2.06$ the dashed curve $a_s=-1.96,\ a_t=-2.06$,
the dash-dot curve $a_s=-1.15,\ a_t=-1.81$ and the dotted
curve $a_s=-0.60,\ a_t=-2.06$ fm. \label{spin}}
\end{figure}

We have seen that it is difficult to determine the  scattering 
lengths individually. It is possible to separate the singlet and 
triplet states with the use of spin information from the deuteron 
initial state or the photon final state.  
Figure \ref{spin} shows the spectra for the six possible
combinations of these spin projections for four selected
parameter sets.  While the overall normalization of these
curves is arbitrary, the relative normalization among them
is correct. The upper left hand curve (for $S_z=+1$
and right circularly polarized photon) shows the maximum
sensitivity since it corresponds to the case where the singlet and 
triplet final states are coherent in a destructive manner.  The 
disadvantage is that this same cancellation leads to a 
small rate if the singlet and triplet scatterings are not too 
dissimilar, as can be expected from SU(6) symmetry.  However, this 
small quantity is a direct measure of the {\em difference between 
the singlet and triplet lengths.} 
Note that the spectra from an 
initial state $S_z=0$ are independent of the singlet parameters.

\begin{figure}[p]
\epsfysize=180mm
\epsffile{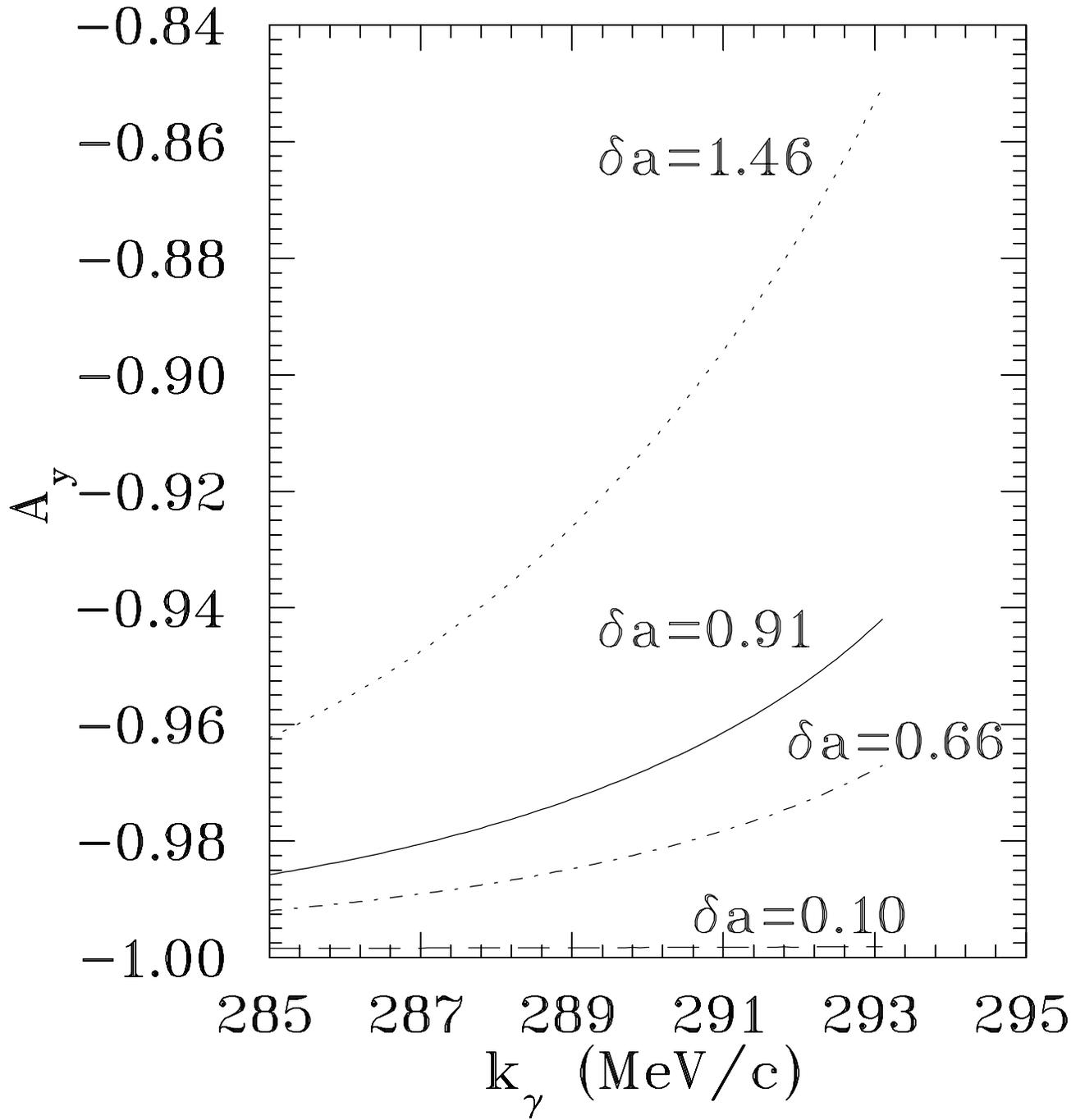}
\caption{ Vector polarization asymmetry for the deuteron
initial spins.  \label{pol1}}
\end{figure}

We can express the results in terms of more conventional
variables. Figure \ref{pol1} illustrates the spectrum expected
from a pure vector polarized deuteron target with the 
asymmetry defined as
\eq
A_y\equiv 
\frac{\Gamma(S_z=+1)-\Gamma(S_z=-1)}
{\Gamma(S_z=+1)+\Gamma(S_z=-1)}.
\qe
We see that a measurement with precision of the order of 2\%
is needed near the peak of the spectrum or 10\% at the
upper part of the spectrum.  What is being searched for
is a deviation of the asymmetry from --1 or, in other
words, a non-null value of the rate from the initial 
state $S_z=+1$.  Such an observation would prove that
the two scattering lengths are unequal. The dashed curve
corresponds to a difference of 0.1 fm, the dash-dot 
curve to 0.66 fm, the solid curve to 0.91 fm and the
dotted curve to 1.46 fm.  This result is almost
completely free from the model dependent effects discussed
previously.  The observation of a difference from --1
in the asymmetry would be a direct indication of a difference 
in the scattering lengths.

\begin{figure}[p]
\epsfysize=180mm
\epsffile{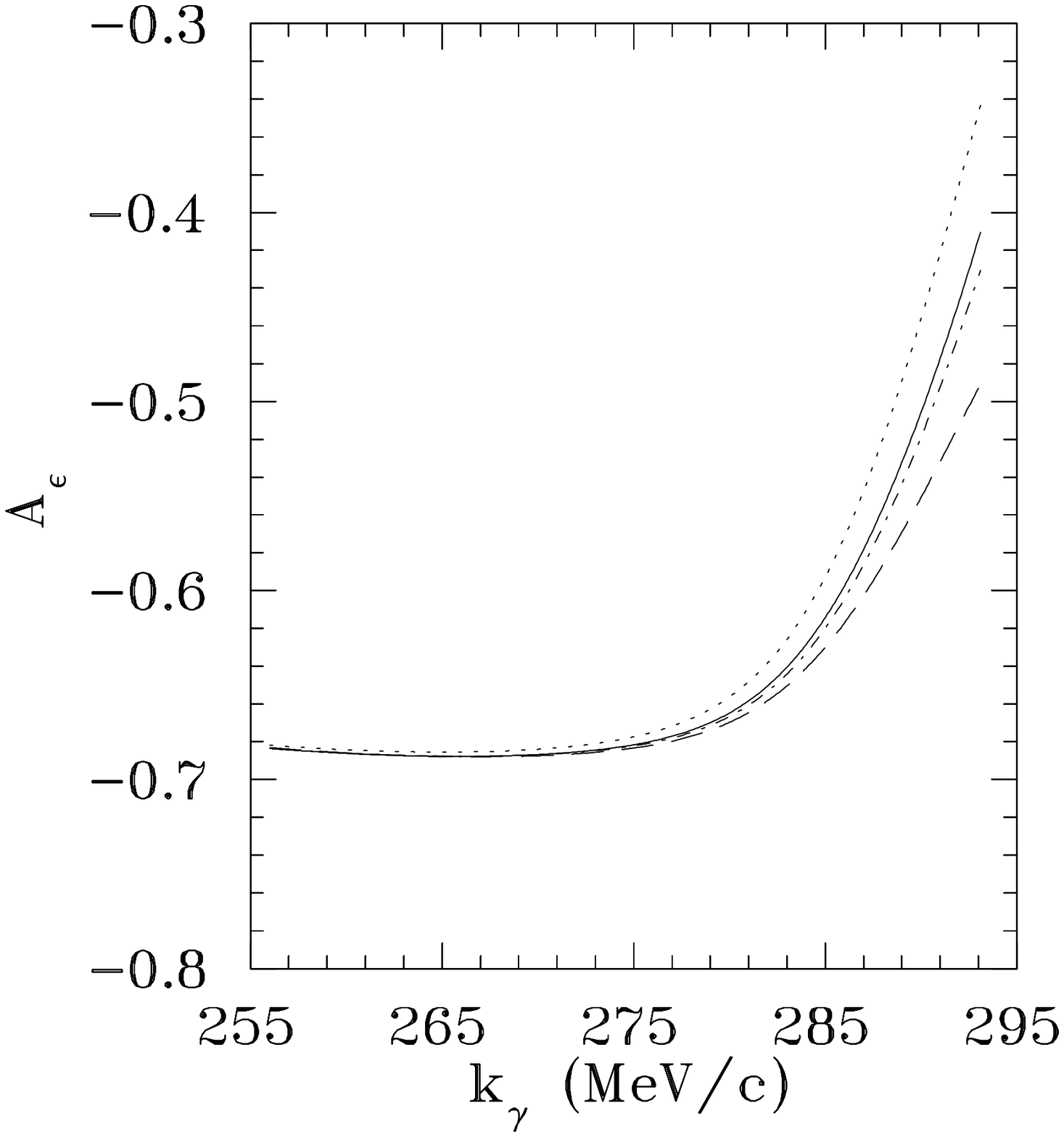}
\caption{Circular polarization of the photon under the same
conditions as the previous figures. \label{cpol}}
\end{figure}

Figure \ref{cpol} shows the spectra expected
from measurements in which the circular polarization of the
final photon is measured. The asymmetry is defined as
\eq
A_{\epsilon}\equiv \frac{\Gamma(\epsilon=+1)-
\Gamma(\epsilon=-1)}
{\Gamma(\epsilon=+1)+\Gamma(\epsilon=-1)}.
\qe
Here again the same order of accuracy is needed to separate
the dotted and dashed curves.

\section{Conclusions}

A formalism has been presented with attention
to the spin degrees of freedom. Due to the difference in shape
of the spectrum for the reaction proceeding from the zero
magnetic quantum number projection of the deuteron from that
with magnetic quantum numbers of $\pm 1$, the singlet-triplet
separation can be made, although the difference is not large
and the triplet state tends to dominate. 

An analysis of the present data which exist for this 
reaction\cite{gall} has been made. Even though the uncertainties 
in the data from this feasibility experiment are
large, some information can be obtained; in particular, separate
numbers can be extracted for the singlet and triplet scattering
lengths (with large overlapping errors). We find for the
singlet scattering length in the range $-0.15$ to $-5.0$ and
the triplet value from $-1.3$ to $-2.65$ (from the 2 fm 
range fit).  

While this precision may seem rather modest,
it is useful to compare with current values in the literature.
Tan\cite{tan} obtained a value (believed to be mostly the
triplet scattering length) of $-2.0\pm 0.5$ fm from an analysis
of data on the reaction $K^-d\rightarrow \pi^-p\Lambda$.
Recent data\cite{balewski} on the reaction $pp\rightarrow 
pK^+\Lambda$, in conjunction with previous elastic 
scattering data\cite{alexander}, lead to a value for the 
spin-averaged scattering length of $-2.0\pm 0.2$ fm\cite{balewski}.
The importance of the influence of the third strongly interacting
particle in the final state is difficult to estimate and the
errors are experimental only.

In a three-body calculation for the hyper-triton bound state  
Miyagawa concludes\cite{miyagawa}
that the best values of the singlet scattering length lie
between $-2.7$ fm and $-2.4$ fm and the triplet between
$-1.6$ fm and $-1.3$ fm, conforming to the ``f'' solution of 
Rijken \ea \cite{rijken}.

An analysis of pseudo data for a $K^-d\rightarrow \Lambda n\gamma$
experiment 
with a cross-comparison of models
leads to an estimate of the model dependence of the order
of $\pm 0.2$ fm. The same type of study gives an estimate of
the error in such an experiment of $\pm 0.3$ fm for the 
triplet scattering length and $\pm 0.8$ fm for the singlet. These
estimates are for measurements without spin information and assume 
an uncertainty of 3\% for the maximum rate in the spectrum. 

It was shown that spin information in either the initial
or final state would be valuable in separating the scattering
lengths.  If it were possible to perform the capture experiment
on a deuteron target of purely magnetic quantum number zero
the triplet scattering length alone would be measured.

The use of a polarized deuteron target would also allow the
separation since the singlet and triplet states interfere
destructively or constructively according to the relative
alignment of the deuteron spin along or against the direction 
of the photon.

Since the previous feasibility study showed that the measurement
was possible but very difficult, one might ask if the increased
difficulty due to the requirement of polarizing the deuteron
might not make it impossible.  Certainly it adds another constraint,
especially if it is necessary to add high-Z material which would
preferentially capture the Kaons.  In this regard we mention a
possible polarized deuteron target which is made from a 
hydrogen-deuteron molecular system\cite{hd1,hd2}.  It has a high density
and contains no heavy materials.

For a measurement of the  circular polarization of the photon, 
the spectra show a significant sensitivity of a similar type.

This work was supported by the U. S. Department of Energy and the 
National Science Foundation.  HKH thanks the Korean Scientific and
Engineering Foundation for their support. The research of BFG
was supported by the Department of Energy (LA-UR-99-6261).

\clearpage

\clearpage

\begin{thebibliography}{300} 

\bibitem{pruhonice} S. A. Coon, H. K. Han, J. Carlson, 
B. F. Gibson, Proceedings of ``Mesons and Light Nuclei 98'', 
p. 407, World Scientific, Singapore

\bibitem{alexander} See Ref. \cite{rijken} for a set of references
to the data.

\bibitem{rijken} Th. A. Rijken, V. G. J. Stoks and Y. Yamamoto,
\prc {\bf 59}, 21(1999)

\bibitem{reuber} A. Reuber, K. Holinde and J. Speth, 
\nucp {\bf A57}, 543(1994)

\bibitem{crowe} R. H. Phillips and K. M. Crowe, \pr {\bf 96},
484(1954)

\bibitem{mcvoy} K. McVoy, \pr {\bf 121}, 1401(1961)

\bibitem{bander} M. Bander, \pr {\bf 134}, B1052(1964)

\bibitem{hadock} R. W. Salter, R. P. Hadock, Z. Zeller, D. R. 
Nygren and J. B. Czirr, \nucp {\bf A254}, 241(1975)

\bibitem{ggs} W. R. Gibbs, B. F. Gibson and G. Stephenson,
\prc {\bf 11}, 90(1975)

\bibitem{psi} O. Schori, B. Gabioud, C. Joseph, J. P. Perroud,
D. R\"uegger, M. T. Tran, P. Tru\"ol, E. Winkelmann and
W. Dahme, \pr C {\bf 35}, 2252(1987); B. Gabioud \ea, \nucp
{\bf A420}, 9(1981)

\bibitem{torrow} W. Tornow \ea, \nucp {\bf A631}, 421c(1998);
C. R. Howell \ea, \pl B 444, 252(1998)

\bibitem{gibson} B. F. Gibson, G. J. Stephenson, 
V. R. Brown, and M. S. Weiss, BNL Report 18335, 1973

\bibitem{workman} R. L. Workman and H. W. Fearing, \prc {\bf 41}, 
1688(1990)

\bibitem{akhiezer} A. I. Akhiezer, G. I Gakh, A. P. Rekalo and
M. P. Rekalo, {\it Yad. Fiz.} {\bf 27}, 214(1978) 
[{\it Sov. J. Nucl. Phys.} {\bf 27}, 115(1978) 

\bibitem{cotanch} R. Williams, C.-R. Ji and S. R. Cotanch, 
\prc 41, 1449(1990); \prc 43, 452(1991)

\bibitem{gall} K. P. Gall \ea, \prc {\bf 42}, R475(1990)

\bibitem{balewski} J. T. Balewski \ea, {\it Eur. Phys. J.} 
A2, 99(1998); \pl B 420, 211(1998)

\bibitem{williams} W. S. C. Williams, ``An Introduction to
Elementary Particles'', First Edition, Academic Press,
New York and London, 1961, p. 374

\bibitem{rosa} T. E. O. Ericson and M. Rosa-Clot, \nucp 
{\bf A405}, 497(1983)

\bibitem{fg}J. L. Friar, B. F. Gibson and G. L. Payne, 
\pr {\bf C30}, 1084(1984) 

\bibitem{ballot} J. L. Ballot and M. R. Robilotta, \pr
{\bf C45}, 986(1992); \pr {\bf C45}, 990(1992);
J. L. Ballot, A. M. Eir\'{o} and M. R. Robilotta, \pr 
{\bf C40}, 1459(1989)

\bibitem{jost} R. Jost, {\it Helv. Phys. Acta.} {\bf 20}, 256(1947)

\bibitem{stoks} V. Stoks, Private Communication

\bibitem{tan} Hai Ho Tan, \prl {\bf 23}, 395(1969)

\bibitem{miyagawa} K. Miyagawa, Proceedings of the First Asian
Few Body Conference

\bibitem{hd1} A. Honig, Q. Fan, X. Wei, A. M. Sandorfi and C. S.
Whisnant, ``7th Workshop on Polarized Target Materials and 
Techniques'', Bad Honnef, Germany, June 20-22, 1994. Nucl.
Instrum. and Meth. A356, 39(1995)
 
\bibitem{hd2} A. Honig, Q. Fan, X. Wei, Breuer, J. P. Didelez,
M. Rigney, M. Lowry, A. M. Sandorfi, A. Lewis and C. S. Whisnant,
``Twelfth Intl. Symp. on High Energy Spin Physics'', Amsterdam,
Sept. 1996, "Spin96" Conf. Proceedings, World Scientific 1997, 
p. 365 
 

\end{thebibliography}
\end{document}